\documentclass[epsfig,twocolumn,showpacs,preprintnumbers,amsmath,amssymb]{revtex4}
\usepackage{graphicx}% Include figure files
\usepackage{dcolumn}% Align table columns on decimal point
\usepackage{bm}% bold math

\begin{document}

%\preprint{ROUGH DRAFT: NOT FOR DISTRIBUTION}

\title{Properties of the superconducting state in a two-band model}
\author{E.J. Nicol}
\affiliation{Department of Physics, University of Guelph,
Guelph, Ontario, N1G 2W1, Canada}
\author{J.P. Carbotte}
\affiliation{Department of Physics and Astronomy, McMaster University,
Hamilton, Ontario, L8S 4M1, Canada}
\date{\today}

\begin{abstract}
Eliashberg theory is used to investigate the range of thermodynamic properties
possible within a two-band model for s-wave superconductivity
and to identify signatures of its two-band nature. 
We emphasize dimensionless
BCS ratios (those for the energy gaps, the specific heat jump and the negative
of its slope
near $T_c$, the thermodynamic critical field $H_c(0)$,
and the normalized slopes of the critical field and the penetration depth
near $T_c$), which are no longer universal
even in weak coupling. We also give results for temperature-dependent
quantities, such as the penetration depth and the energy gap.
 Results are presented both for microscopic parameters
appropriate to MgB$_2$ and for variations away from these.
Strong coupling corrections are identified and found to be significant.
Analytic formulas are provided which show the role played by
the anisotropy in coupling in
some special limits.
Particular emphasis is placed on small 
interband coupling and on the opposite limit of no diagonal coupling.
The effect of impurity scattering is considered, particularly for
the interband case.
\end{abstract}

\maketitle
% body of paper here

\section{Introduction}

The properties of the superconducting state of conventional, single-band,
electron-phonon superconductors differ markedly from BCS predictions.\cite{carbotte}
However, they are well-described within isotropic Eliashberg theory
with a single electron-phonon spectral density $\alpha^2F(\omega)$
for the average interaction over the Fermi surface. This function
is accurately known from inversion of tunneling data.\cite{rowell}
 In many cases,
the $\alpha^2F(\omega)$ has also been calculated from first principle
electronic band structure calculations extended to include the
electron-phonon interaction, sometimes with the phonons taken
directly from inelastic neutron scattering measurements. In many cases,
such results agree very well with the corresponding tunneling data.
While it is to be noted that, in principle, the electron-phonon spectral
density for the various electrons on the Fermi surface is anisotropic
leading to energy gap anisotropy\cite{marsiglio,daams,tomlinson,leung}, 
this feature often does not play
a prominent role because, in many instances, the electronic mean free
path is much smaller than the coherence length. In such circumstances,
a Fermi surface average of the electron-phonon spectral density can
be used. Nevertheless, corrections due to gap anisotropy have
been identified and studied in the past\cite{webber}, often, but not always,
in a separable anisotropic  model.\cite{zarate}

The history of two-band superconductivity\cite{suhl,others,rainer,entel}
 and of MgB$_2$ (with $T_c\simeq 39$ K\cite{akimitsu})
in particular\cite{liu,choi,choinature,dolgovpen,golubov,goliop,an,mazin,mazinrev}
is somewhat different. To our knowledge, as yet, there exists no inversion\cite{yanson}
of tunneling data from which the electron-phonon interaction is determined.
In fact, it has been noted\cite{dolgov}
 that this may well never be possible in MgB$_2$
because of its two-band nature which requires a microscopic description
in terms of four separate electron-phonon spectral functions $\alpha^2_{ij}
F(\omega)$, where $i=\sigma,\pi$ (or 1, 2), with the two-dimensional $\sigma$ band
having the largest electron-phonon coupling. The three-dimensional $\pi$
band on its own would have a smaller value of $T_c$, the critical temperature,
although it has a  higher value of the electron density of states
at the Fermi energy.

In the absence of tunneling data giving reliable information on the fundamental
kernels entering the two-band Eliashberg equations, first principle
band structure calculations of $\alpha^2_{ij}F(\omega)$ in MgB$_2$
have been used to compute superconducting properties (for example, \cite{goliop,dolgovpen,choinature,golubov}).
 To do this, it is also
necessary to know the Coulomb pseudopotential 
repulsions $\mu^*_{ij}$ which are different
for various indices ($i,j$), but these have also been
calculated. Good agreement with experiment is obtained in this way for the
properties considered so far, more explicitly, the specific heat\cite{choinature,junod}, the
penetration depth\cite{dolgovpen,moca},
and the anisotropy in the two gaps, as well as their temperature dependence. 
For the penetration depth, impurity scattering can be 
important, and in and out of plane orientation of the magnetic field are
different\cite{dolgovpen}.

In this paper, we use the band theory information on $\alpha^2_{ij}F(\omega)$
and $\mu^*_{ij}$ in MgB$_2$ to calculate  
the critical temperature, the energy gap with its anisotropy and temperature
dependence, and other thermodynamic properties, as well
as the penetration depth, giving particular emphasis to strong coupling
corrections. Further to our discussion
of MgB$_2$, we provide a full listing of calculated dimensionless BCS ratios,
now modified by both the anisotropy and the strong coupling effects in
MgB$_2$, and make comparison with experiment.
We also consider effects of variations in microscopic
parameters away from those of MgB$_2$, as well as impurity scattering
- intraband and interband. To this end,
we reduce the two-band Eliashberg equations, which fully account for 
retardation, in the two-square-well approximation (also called
the $\lambda^{\theta\theta}$ model). This leads to simple
{\it renormalized} BCS (RBCS) forms which, when compared to our full numerical
Eliashberg results, allow us to identify the strong coupling corrections
which we find to be significant even for MgB$_2$. 

When considering variations in microscopic parameters away from those of
MgB$_2$, we place particular emphasis on two limiting cases: the
limit of small interband coupling and the opposite case, when the 
intraband coupling is zero and the superconductivity is due to the interband
coupling alone, a case discussed in the early work of Shul {\it et al.}\cite{suhl}. We also consider the special case when the intraband coupling in
the second band is repulsive. The limit of small interband coupling
is particularly interesting because it allows us to understand 
how the offdiagonal terms lead to the integration of otherwise two
completely independent and non-communicating superconducting bands
with separate transition temperatures $T_{ci}$. In this regard, we find
that $\alpha^2_{12}F(\omega)$ and $\alpha^2_{21}F(\omega)$ behave
very differently with 21 the most effective variable at integrating the
two systems and 12
the most effective at changing the critical
temperature. The presence of the offdiagonal interactions rapidly smear
out the features of the second transition at $T_{c2}$, {\it i.e}, the one with the smaller
of the two $T_{ci}$ values. More specifically, surprisingly small values of the
mass renormalization parameter $\lambda_{21}$, as compared
with $\lambda_{11}$ and $\lambda_{22}$, have a large effect on the region
of $T_{c2}$. We also find that relatively modest values of
the interband impurity 
scattering rates lead to the significant integration of the two
bands. Even when the bands are well-integrated, in the sense that little
trace of a second sharp transition at $T_{c2}$ remains, there still exist
important modifications of the usual one-band BCS results because of the
two distinct bands. As an example, the BCS dimensionless universal ratios
now depend on  the ratio of the electronic density of states at the Fermi
energy  of the two bands. Simple analytic expressions for these ratios are
derived, which provide
insight into the physics underlying two-band superconductivity 
and guidance as to how these results are to be interpreted.

In section II, we give the two-band Eliashberg equations and provide their
reduction in the $\lambda^{\theta\theta}$ approximation which is needed to
identify strong coupling corrections to renormalized two-band
BCS (RBCS). Section III deals with the
dependence of $T_c$ on microscopic parameters, {\it i.e}, on the
electron-phonon interaction as well as on impurities. Intra- and interband 
quantities are both of interest. We consider the modifications
of the dimensionless BCS
ratios in the $\lambda^{\theta\theta}$ model, as well as, the zero temperature
value of the two gaps and their anisotropy. MgB$_2$ 
is considered in section IV.
The issue of strong coupling corrections in
MgB$_2$, and more generally in other related systems, is discussed.
The limit of small interband
electron-phonon coupling is considered in section V. We study, in 
particular, how the two otherwise separate bands become integrated when this
interaction is switched on. The effect of interband impurity scattering
is also considered in the same context as it exhibits
analogous behaviour to the case of the offdiagonal
electron-phonon coupling. In section VI, we deal briefly
with the less realistic case of zero intraband electron-phonon coupling,
 where the superconductivity is due only to the interband piece. Conclusions
are found in section VII.

Finally, in light of the recent
developments in other areas of superconductivity and correlated
electrons, we wish to emphasize that our use of the term
``gap anisotropy'' here is
in reference to the difference in the magnitudes of the two gaps, each of
which
are isotropic s-wave in this work, and hence 
does not refer to a momentum-dependent order parameter.
Likewise, ``strong coupling'' refers
to the traditional meaning of strong electron-phonon coupling and is
not an allusion to strong interband coupling. 

\section{Theory}

The isotropic (within a band) Eliashberg equations generalized to two bands
$(i=1,2)$ are written on the imaginary axis 
as\cite{entel,carbotte,marsiglio,daams}:
\begin{eqnarray}
\Delta_i(i\omega_n)Z_i(i\omega_n) &=& \pi T\sum_m\sum_j[\lambda_{ij}(i\omega_m-i\omega_n)
\nonumber\\
&-&\mu^*_{ij}(\omega_c)\theta(\omega_c-|\omega_m|)]
\frac{\Delta_j(i\omega_m)}{\sqrt{\omega_m^2+\Delta_j^2(i\omega_m)}}\nonumber\\
&+& \pi\sum_j(t^+_{ij}-t^-_{ij})\frac{\Delta_j(i\omega_n)}{\sqrt{\omega_n^2+\Delta_j^2(i\omega_n)}}\label{eq:Del}
\end{eqnarray}
and
\begin{eqnarray}
Z_i(i\omega_n) &=& 1+\frac{\pi T}{\omega_n}\sum_m\sum_j
\lambda_{ij}(i\omega_m-i\omega_n)\nonumber\\
&\times&
\frac{\omega_m}{\sqrt{\omega_m^2+\Delta_j^2(i\omega_m)}}\nonumber\\
&+& \pi\sum_j(t^+_{ij}+t^-_{ij})\frac{\omega_n}{\sqrt{\omega_n^2+\Delta_j^2(i\omega_n)}},
\label{eq:Z}
\end{eqnarray}
where $t^+_{ij}=1/(2\pi\tau^+_{ij})$ 
and  $t^-_{ij}=1/(2\pi\tau ^-_{ij})$ are the
ordinary and paramagnetic impurity scattering rates, respectively,
and
\begin{equation}
\lambda_{ij}(i\omega_m-i\omega_n)\equiv 2\int^\infty_0\frac{\Omega\alpha^2F_{ij}(\Omega)}{
\Omega^2+(\omega_n-\omega_m)^2}d\Omega.
\end{equation}
Eq.~(\ref{eq:Del}) gives the gap $\Delta_i(i\omega_n)$ and Eq.~(\ref{eq:Z}),
the renormalization $Z_i(i\omega_n)$ at the $n$'th Matsubara frequency $i\omega_n$,
with $\omega_n=(2n-1)\pi T$. Here, $T$ is temperature and $n=0,\pm 1, \pm 2,\cdots$. The electron-phonon kernels are $\alpha^2_{ij}F(\Omega)$ as a function
of phonon energy $\Omega$ and the Coulomb repulsions are $\mu^*_{ij}$,
with a high energy cutoff $\omega_c$ needed for convergence and
usually taken to be about six to ten 
times the maximum phonon frequency. 
For the specific case of MgB$_2$, these may be found
in \cite{goliop}.
The diagonal intraband
elements of the electron-phonon interaction are largest, in the case
of MgB$_2$, while the offdiagonal
elements describing interband scattering are smaller, but still substantial.

In what is called the two-square-well approximation or $\lambda^{\theta\theta}$
model\cite{rainer,allen,am}, we use in Eq.~(\ref{eq:Del}):
\begin{eqnarray}
\lambda_{ij}(i\omega_m-i\omega_n) &=& \lambda_{ij}, {\,\rm for\,} {\rm both\,} |\omega_n|,|\omega_m|<\omega_\circ\nonumber\\
&=&0, {\quad\rm otherwise,}
\end{eqnarray}
where
\begin{equation}
\lambda_{ij}(m=n)=\lambda_{ij}(0)\equiv\lambda_{ij}=2\int^\infty_0
\frac{\alpha^2F_{ij}(\Omega)}{\Omega}d\Omega.
\end{equation}
Neglecting the gap in the denominator on the right-hand side of 
Eq.~(\ref{eq:Z}) for $Z$, we further approximate 
(see Ref.~\cite{am} for details)
\begin{equation}
Z_i(i\omega_n)=1+\sum_j\lambda_{ij}.
\end{equation}
This result may now be used in Eq.~(\ref{eq:Del})
to obtain
\begin{eqnarray}
\Delta_i(i\omega_n) &=& \Delta_i(T),\, |\omega_n|<\omega_\circ\nonumber\\
            &=& 0,\, |\omega_n|>\omega_\circ,
\end{eqnarray}
where
\begin{equation}
\Delta_i(T)=\frac{\pi T}{Z_i}\sum_{m,|\omega_m|<\omega_\circ}\sum_j
\frac{\Delta_j(T)}{\sqrt{\omega_m^2+\Delta_j^2}}[\lambda_{ij}-\mu^*_{ij}],
\label{eq:rbcsdel}
\end{equation}
where $\omega_\circ$ represents either the Debye frequency or some other
characteristic energy scale representing the phonons in the system, at
most the maximum phonon energy.  Detailed justification of using a single
cutoff is found in Ref.~\cite{rainer}.
These results are used to derived various quantities within the $\lambda^{\theta\theta}$ model, which we will call renormalized BCS or RBCS. We also
solve the full Eliashberg equations for typical strong coupling parameters
and for the case of MgB$_2$, and in order to connect to the language most
appropriate for this purpose, the measure of the characteristic
boson frequency, $\omega_{\ln}$, is defined to be:
\begin{equation}
\omega_{\ln} = {\rm exp}\bigg[\frac{2}{\lambda_{11}}\int_0^\infty\ln(\omega)
\frac{\alpha^2F_{11}(\omega)}{\omega}d\omega\biggr].
\end{equation}
This is reasonable for our case here as the $\omega_{\ln}$
calculated for the different $\alpha_{ij}^2F(\omega)$ spectra of MgB$_2$
are almost the same and other spectra used in this paper will
 have the same
frequency distribution in each channel only scaled in magnitude. In general,
 this definition should be reasonably robust
as, unless $\lambda_{22}$, $\lambda_{12}$, and $\lambda_{21}$,
are large, the first channel $\lambda_{11}$ should dominate the strong coupling
effects.

\section{BCS ratios: the $\lambda^{\theta\theta}$ model and strong coupling}
\subsection{Critical Temperature: $T_c$}

The critical temperature that results from the renormalized BCS equation
(\ref{eq:rbcsdel}) of the two-square-well approximation, takes the form
\begin{equation}
A=\ln\biggl(\frac{1.13\hbar\omega_\circ}{k_BT_c}\biggr),
\end{equation}
or
\begin{equation}
k_BT_c= 1.13\hbar\omega_\circ e^{-A},
\label{eq:Tc}
\end{equation}
where
\begin{equation}
A=\frac{\bar\lambda_{11}+\bar\lambda_{22}-\sqrt{(\bar\lambda_{11}
-\bar\lambda_{22})^2
+4\bar\lambda_{12}\bar\lambda_{21}}}
{2(\bar\lambda_{11}\bar\lambda_{22}-\bar\lambda_{12}\bar\lambda_{21})}
\label{eq:A}
\end{equation}
and
\begin{eqnarray}
\bar\lambda_{11} = \frac{\lambda_{11}-\mu^*_{11}}{1+\lambda_{11}+\lambda_{12}},&&
\bar\lambda_{12} = \frac{\lambda_{12}-\mu^*_{12}}{1+\lambda_{11}+\lambda_{12}},\nonumber\\
\bar\lambda_{22} = \frac{\lambda_{22}-\mu^*_{22}}{1+\lambda_{22}+\lambda_{21}},&&
\bar\lambda_{21} = \frac{\lambda_{21}-\mu^*_{21}}{1+\lambda_{22}+\lambda_{21}}.
\label{eq:barlam}
\end{eqnarray}
With no impurities and for one band ($\lambda_{12}=\lambda_{21}=\lambda_{22}=0$)
\begin{equation}
k_BT_{c}^{00}=1.13\omega_\circ e^{-1/\bar\lambda_{11}}.
\label{eq:Tc00}
\end{equation}
Here we will be interested only in the ratio of $T_c$ (Eq.~(\ref{eq:Tc}))
to $T_c^{00}$ (Eq.~(\ref{eq:Tc00})) and so the cutoff $\omega_\circ$
cancels, and the issue of the best choice for this quantity does not
enter (see Allen and Dynes\cite{allen}).
\begin{figure}[ht]
\begin{picture}(250,200)
\leavevmode\centering\includegraphics{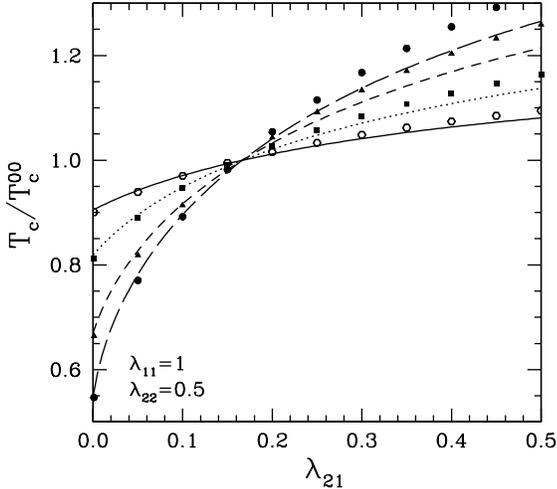}
\end{picture}
%\vskip 60pt
%\begin{figure}
%\includegraphics[clip,width=0.45\textwidth]{tc1.eps}
%\includegraphics{tc1.eps}
\caption{Ratio of $T_c$ to the pure, one-band $T_c^{00}$  as a function of
$\lambda_{21}$ for varying $\lambda_{12}$: 0.6 (long-dashed), 0.4 (short-dashed),
0.2 (dotted), and 0.1 (solid). Here, $\lambda_{11}=1$ and $\lambda_{22}=0.5$.
Strong coupling Eliashberg calculations are given for comparison for the same
parameters and are shown as the points with $\lambda_{12}$: 0.6 (solid circles), 0.4 (solid triangles), 0.2 (solid squares), and 0.1 (open circles).}
\label{fig1}
\end{figure}
Results for $T_c/T_c^{00}$ based on Eqs.~(\ref{eq:Tc}-\ref{eq:Tc00})
as a function of $\lambda_{21}$ for various values of $\lambda_{12}$
are shown in Fig.~\ref{fig1}, where they are compared with results of
complete numerical evaluation of the two-band Eliashberg equations
(\ref{eq:Del}) and (\ref{eq:Z}). A Lorentzian
model for the spectral densities $\alpha^2_{ij}F(\omega)$
is used with zero Coulomb pseudopotential $\mu^*_{ij}$ for simplicity.
Specifically, we use a truncated
Lorentzian spectral density, which is defined in
Ref.~\cite{nicol},
centered around 50 meV with width 5 meV, truncated by 50 meV
to either side of the central point. The $\omega_{\ln}$
for this spectrum is 44.6 meV. This spectral density is scaled
in each of the four channels to give
$\lambda_{11}=1$, $\lambda_{22}=0.5$, and the range of values
of $\lambda_{12}$ and $\lambda_{21}$ as required for the figure.
The curves, which are labelled in the figure caption, 
are for the renormalized BCS
calculations and the corresponding 
Eliashberg calculations are presented as points.
We note that for small values of $\lambda_{21}$
agreement between the $\lambda^{\theta\theta}$ results and full Eliashberg
is excellent. The agreement is somewhat less good around $\lambda_{21}=0.5$
but still acceptable. An interesting point to note about this figure is that
the effect on $T_c$ of $\lambda_{21}$ and $\lambda_{12}$ are quite different.
As $\lambda_{21}$ increases for fixed $\lambda_{12}$, $T_c$ increases. On
the other hand, for small but fixed $\lambda_{21}$, increasing $\lambda_{12}$
decreases $T_c$, while the opposite behaviour is found to hold for values
of $\lambda_{21}$ bigger than approximately 0.16. This behaviour is different
from that expected in non-renormalized 
BCS theory where it is known that increasing 
the offdiagonal coupling from zero to some finite value always
increases $T_c$ whatever its sign. Expanding Eq.~(\ref{eq:A}) under the 
assumption that the offdiagonal elements are small 
as compared with  the diagonal
ones ($\bar\lambda_{12},
\bar\lambda_{21}\ll \bar\lambda_{11}-\bar\lambda_{22},\bar\lambda_{22}$) gives
\begin{equation}
A\simeq \frac{1}{\bar\lambda_{11}}\Biggl[1-\frac{\bar\lambda_{12}\bar\lambda_{21}}{\bar\lambda_{22}}
\biggl\{\frac{1}{\bar\lambda_{11}-\bar\lambda_{22}}-\frac{1}{\bar\lambda_{11}}\biggr\}\Biggr].
\end{equation}
In BCS theory, the $\bar\lambda_{ij}$ would not be renormalized
as in Eq.~(\ref{eq:barlam}). Since the term in curly brackets is positive,
$A$ decreases with the product of $\bar\lambda_{12}\bar\lambda_{21}$
and hence $T_c$ increases. But in our case, the multiplying term $1/\bar\lambda_{11}
\equiv (1+\lambda_{11}+\lambda_{12})/\lambda_{11}$ contains $\lambda_{12}$
in leading order and this factor on its own increases $A$ and therefore
decreases the critical temperature. These expectations are confirmed in
our full Eliashberg numerical work
and are not captured in 
other BCS works (for example \cite{kita,mishonov}). 
It is clear then, that in our theory,
$\lambda_{12}$ and $\lambda_{21}$ do not enter the equation for $T_c$
in the same way because $\lambda_{12}$ provides a direct mass renormalization
to the major interaction term $\lambda_{11}$. If mass renormalization
is ignored, as in BCS theory, this asymmetry no longer arises.
The work by Mitrovi\'c\cite{mitrovic} on functional derivatives
finds $\delta T_c/\delta \alpha^2F_{21}(\omega)$ to be positive
and the one for 12 to be negative, which conforms with our results.
We note here that the disparity between $\lambda_{12}$ and $\lambda_{21}$,
which will in turn affect the $T_c$ and other properties,
is related to the 
different values of the density of states at the Fermi level $N_i$
in each of the two bands,
{\it i.e.} $\lambda_{12}/\lambda_{21}=N_2/N_1$.

Turning next to the effect of impurities on $T_c$,
 the change  $\Delta T_c = T_c-T_{c0}$
for small impurity scattering can
be written in the $\lambda^{\theta\theta}$ model as:
\begin{equation}
\frac{\Delta T_c}{T_{c0}}=\frac{C^\pm}{\bar\lambda_{11}+\bar\lambda_{22}+
2A(\bar\lambda_{12}\bar\lambda_{21}-\bar\lambda_{11}\bar\lambda_{22})},
\label{eq:tcimp}
\end{equation}
where for ordinary impurities ($C^+$) and magnetic impurities ($C^-$):
\begin{eqnarray}
C^{\pm}&=&-\frac{\pi^2}{4}\{(1-A\bar\lambda_{22})(\rho^{\pm}_{12}\bar\lambda_{11}\mp\rho^{\pm}_{21}\bar\lambda_{12})\nonumber\\
&+&(1-A\bar\lambda_{11})(\rho^{\pm}_{21}\bar\lambda_{22}\mp\rho^{\pm}_{12}\bar\lambda_{21})\nonumber\\
&+&A\bar\lambda_{21}(\bar\lambda_{12}\mp\bar\lambda_{11})\rho^{\pm}_{12}\nonumber\\
&+&A\bar\lambda_{12}(\bar\lambda_{21}\mp\bar\lambda_{22})\rho^{\pm}_{21}\},
\end{eqnarray}
with
\begin{equation}
\rho^{\pm}_{12}=\frac{t^{\pm}_{12}/T_{c0}}{1+\lambda_{11}+\lambda_{12}}, \,\,
\rho^{\pm}_{21}=\frac{t^{\pm}_{21}/T_{c0}}{1+\lambda_{22}+\lambda_{21}}.
\label{eq:rhoimp}
\end{equation} 
These equations have been derived for scattering across the
bands; within the bands, paramagnetic impurities will affect $T_c$
but ordinary, nonmagnetic ones will not.

\begin{figure}[ht]
\begin{picture}(250,200)
\leavevmode\centering\includegraphics{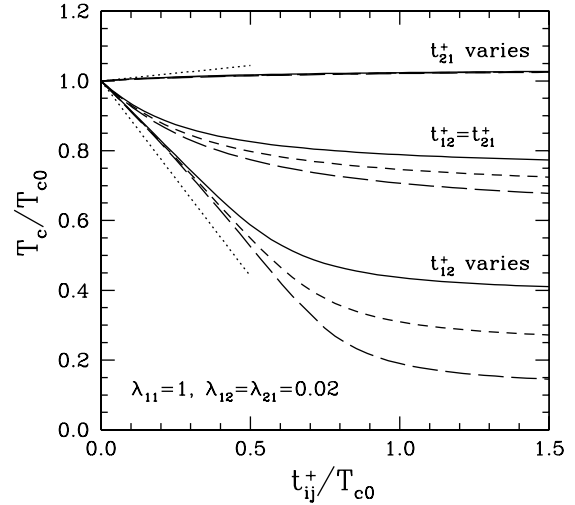}
\end{picture}
%\vskip 60pt
%\begin{figure}
%\includegraphics[clip,width=0.45\textwidth]{tc1.eps}
%\includegraphics{tc1.eps}
\caption{Ratio of $T_c$ with impurity scattering
to that without $T_{c0}$  as a function of
$t^+_{ij}/T_{c0}$ for varying $\lambda_{22}$: 0.5 (solid), 0.4 (short-dashed),
and 0.3 (long-dashed). Here, $\lambda_{11}=1$, $\lambda_{12}=
\lambda_{21}=0.02$. For the lower
three curves $t^+_{21}=0$ and $t^+_{12}$ varies,
and for the upper three curves (which are almost indistinguishable
from each other) it is
the reverse. 
In the middle set of three curves, $t^+_{12}=t^+_{21}$. 
These calculations have been
done with the full Eliashberg equations using a Lorentzian $\alpha^2F(\omega)$
spectrum with $T_{c0}/\omega_{\ln}=0.11$. The dotted lines are from the
evaluation of Eq.~(\ref{eq:tcimp}) for the $\lambda_{22}=0.5$ case
and are for $t^+_{12}=0$ (upper dotted curve) and
$t^+_{21}=0$ (lower dotted curve). (Note that the middle set of curves
are the only physically realizable cases. The others serve to make the
mathematical point that $t^+_{12}$ and $t^+_{21}$ affect $T_c$ quite differently.)}
\label{fig2}
\end{figure}

Results are given in Fig.~\ref{fig2}. Except for the dotted lines,
all curves were obtained from numerical solutions of the linearized
version of the Eliashberg equations (\ref{eq:Del}) and (\ref{eq:Z})
using a Lorentzian model for $\alpha^2_{ij}F(\omega)$. The curves
come in sets of three for $\lambda_{22}=0.5$ (solid curve),
0.4 (dashed) and 0.3 (long-dashed). The other parameters are $\lambda_{11}=1$
and $\lambda_{12}=\lambda_{21}=0.02$ (small interband coupling). The
lower set are for $t^+_{21}=0$ with $t^+_{12}$ varying while
the upper set have $1\leftrightarrow 2$. The middle set have
$t^+_{12}=t^+_{21}$. 
Note that in showing the results when $t^+_{12}$
or $t^+_{21}$ are
varied separately, we are violating a requirement that they must be linked
together by the density of states in the two bands. That is, as required
for the $\lambda_{ij}$'s, likewise the impurity scattering rates must obey
$t^+_{12}/t^+_{21}=N_2/N_1$. Our middle set of curves obey
this constraint, but we have ignored it for the other curves in order
to illustrate the general behaviour of each 
individual type of scattering separately.
As found for the $\lambda_{ij}$'s,
the effect of $t^+_{12}$ and  that of $t^+_{21}$ on $T_c$ are quite different.
The quantity $t^+_{12}$ represents scattering 
 from band 1 to band 2
and leads to pairbreaking much like paramagnetic impurities
in the one-band case. We can see this analytically in the simple case 
of $\lambda_{12}=\lambda_{21}=0$ for which the two bands are decoupled
and the critical temperature is a property of the first band alone.
In this case, Eqs.(\ref{eq:tcimp}-\ref{eq:rhoimp}) reduce to\cite{golubovimp}
\begin{equation}
\frac{\Delta T_c}{T_{c0}}=-\frac{\pi^2}{4}\rho^\pm_{12}
\label{eq:AG}
\end{equation}
for both normal or paramagnetic impurities in the linear approximation for
the
impurity scattering rate. The initial
linear decrease in $T_c$ with increasing
$\rho^+_{12}$ is seen in the lower set of three curves of
Fig.~\ref{fig2}. As $t^+_{12}$ is increased further, higher order
corrections start to be important and the curves show saturation to a value
which is larger, the greater the value of $\lambda_{22}$. Also note
that formula (\ref{eq:AG}) shows that $T_c$ is independent of $\rho^\pm_{21}$.
This expectation is confirmed in the upper set of three curves of Fig.~\ref{fig2},
where  $T_c$ has increased by no more than 3\% for $t^+_{21}/T_{c0}=1.5$.
This small increase is due to the small $\lambda_{12}=\lambda_{21}$ used for
the figure, while in Eq.~(\ref{eq:AG}), we have $\lambda_{12}=\lambda_{21}=0$.
The middle set of curves, which apply for 
$t^+_{12}=t^+_{21}$ and therefore satisfy the constraint imposed by having
chosen $\lambda_{12}=\lambda_{21}=0.02$, exhibits, by comparison to the
other two cases, only a very small region which is linear in impurity
scattering and these curves are intermediate to the other two sets,
as expected. They also saturate at higher values of $T_c$ and we
find that $T_c$ decreases by only 20-30\% for this case, similar to
the observation by Mitrovi\'c who was considering specifically the case of
MgB$_2$\cite{bozaimp}.
Finally, we comment
on the dotted curves which are based on Eqs.~(\ref{eq:tcimp}) to (\ref{eq:rhoimp}) valid in the $\lambda^{\theta\theta}$ model and first order in $t^+_{ij}$.
The lowest curve applies to the $t^+_{21}=0$ case
and the upper one to $t^+_{12}=0$. 
The slopes are in good agreement with the full Eliashberg results over a significant range of
interband impurity scattering $t^+_{ij}$. For the middle set of curves
the linear behaviour applies only comparatively to a rather small region. 
In all cases there still is some difference between 
$\lambda^{\theta\theta}$ results and Eliashberg because of strong coupling
corrections. As previously stated, 
interband impurity scattering in two-band superconductivity
works like paramagnetic impurities in the ordinary one-band case.
 For this latter case, Schachinger, Daams, and Carbotte\cite{ewald}
have found for the specific case of Pb, the classic strong coupling material,
that the $\lambda^{\theta\theta}$ model overestimates the initial slope of the
drop in $T_c$ value, with increasing impurity scattering. The physics is simple.
For strong coupling, $2\Delta/k_BT_c$ is larger than its BCS value
{\it i.e.}, the gap is bigger than expected on the basis of its $T_c$.
This is because  as $T$ is increased, that part of the
 inelastic scattering which corresponds to the real (as opposed to
virtual) processes, which
are pairbreaking, increases
and $T_c$ is reduced below the value it would be without. As a result,
the initial drop in $T_c$ value with increasing impurity content is not
as large in strong as in weak coupling because the system
has a larger gap which is more robust against
impurities. The same applies to interband scattering
 in a two-band superconductor. The initial
slope of the drop is faster in the $\lambda^{\theta\theta}$
model than in Eliashberg, as most recently
shown by Mitrovi\'c\cite{bozaimp}, who has commented on 
prior work by Golubov and Mazin\cite{golubovimp}, where only
unrenormalized BCS results were given and the drop
in $T_c$ was even faster. Mitrovi\'c also presents functional
derivatives  for ordinary impurities\cite{bozaimp} and his findings compliment
our calculations here. In addition, as low frequency phonons
act like ordinary impurities, the previous work by Mitrovi\'c on functional
derivatives\cite{mitrovic}
 for the electron-phonon spectral functions also confirms 
our impurity results by comparison with the behaviour of the low frequency 
part of the functional derivatives for 12 versus 21.

Finally, it has been of some interest amongst experimentalists, looking
at novel superconductors, to know
the outcome of having a repulsive interaction in the second band ({\it i.e.}
$\lambda_{22}<0$). As will be seen in the next section, a second energy
gap is still induced in this case due to the interband coupling,
however, a signature of this repulsive band would exist in the case
of impurity scattering, as strong interband scattering of sufficient
strength could drive the $T_c$ to
zero\cite{kita}.

\subsection{Energy Gaps and Gap Ratios}

We turn next to the consideration of the energy gaps.
The transcendental equation for $u\equiv \Delta_2/\Delta_1$
 at $T=0$ in the $\lambda^{\theta\theta}$ model is:
\begin{equation}
\bar\lambda_{12}u-\frac{\bar\lambda_{21}}{u}+(\bar\lambda_{11}\bar\lambda_{22}
-\bar\lambda_{21}\bar\lambda_{12})\ln u=\bar\lambda_{22}-\bar\lambda_{11},
\label{eq:u}
\end{equation}
from which the gap ratio for the larger gap $\Delta_1$ may be found:
\begin{equation}
\ln\biggl(\frac{1.13\Delta_1}{2k_BT_c}\biggr)=A-\biggl[\frac{1+\bar\lambda_{12}
u\ln u}{\bar\lambda_{11}+\bar\lambda_{12}u}\biggr].
\label{eq:gratio}
\end{equation}
The solution for the gap ratio $2\Delta_1/k_BT_c$ can be corrected for
strong coupling effects by multiplying by a factor $\eta_\Delta$ 
in the
denominator of the logarithm of Eq.~(\ref{eq:gratio}) with\cite{mzc}:
\begin{equation}
\eta_\Delta = 1+12.5\biggl(\frac{T_c}{\omega_{\ln}}\biggr)^2\ln\biggl(\frac{\omega_{\ln}}
{2T_c}\biggr).
\label{eq:etaDel}
\end{equation}
As long as $\lambda_{11}$ is large and $\lambda_{22}$, $\lambda_{12}$,
and $\lambda_{21}$ are small, one needs only to correct the first
channel for strong coupling effects. Otherwise additional corrections for
the other channels may exist but there would be no merit in such
complexity of including these corrections over doing the full numerical
calculations with the Eliashberg equations. It is expected that in
real systems, $\lambda_{11}$ is large relative to the other parameters
and hence dominates the strong coupling aspect of the result.
However, when the offdiagonal couplings are significant, the strong coupling
corrections of the first channel can affect the second.

\begin{figure}[ht]
\begin{picture}(250,200)
\leavevmode\centering\includegraphics{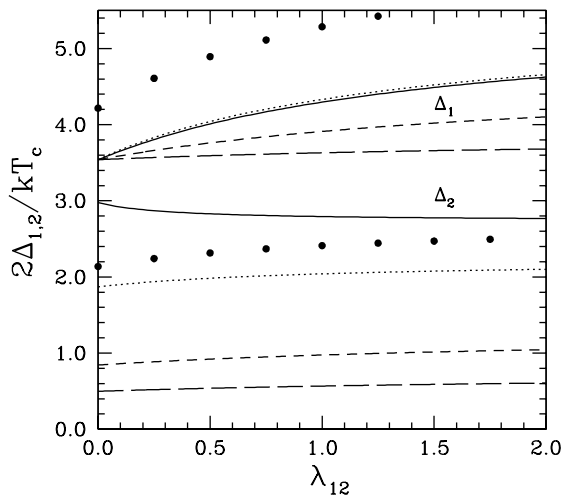}
\end{picture}
%\vskip 70pt
%\begin{figure}
%\includegraphics[clip,width=0.45\textwidth]{tc1.eps}
%\includegraphics{tc1.eps}
\caption{Gap ratios for the upper ($2\Delta_1/k_BT_c$)
and lower gap ($2\Delta_2/k_BT_c$) as a function of $\lambda_{12}$
for varying $\lambda_{22}$: 0.5 (solid), 0.1 (dotted),
-0.5 (short-dashed), and -1 (long-dashed). 
Here, $\lambda_{11}=1$, $\lambda_{21}=0.3$. These calculations are
done using the RBCS formulas (\ref{eq:u}-\ref{eq:gratio})
in the text, the solid dots show 
Eliashberg calculations for the same set of parameters with $\lambda_{22}=0.1$
(for comparison with the dotted curve).
Strong coupling corrections are significant 
and the rest of the curves in this figure would also be
modified by strong coupling, much of this can be captured by the
strong coupling correction formula given in the text. (Note that
as $\lambda_{22}$ and $\lambda_{21}$ are finite, the points for 
$\lambda_{12}=0$ are not physically realizable.)}
\label{fig3}
\end{figure}
Our first set of results for the two energy
gaps
 is given in Fig.~\ref{fig3}.
The lines are based on the simpler equations (\ref{eq:u}) and (\ref{eq:gratio}),
and the solid dots are for the results of full Eliashberg solutions
on the imaginary axis and
analytically continued with Pad\'e approximates\cite{carbotte}
to the real axis, where the gap is 
determined by $\Delta_0=\Delta(\omega=\Delta_0)$\cite{carbotte}. 
For clarity in the figure,
only one such set of results is shown for the case of $\lambda_{22}=0.1$.
While magnitudes differ considerably between the renormalized BCS and
strong coupling (comparing solid dots with the dotted curves), the general
trends are the same. Specifically in Fig.~\ref{fig3}, $\lambda_{12}$ is varied
with $\lambda_{11}=1$, $\lambda_{21}=0.3$, and $\lambda_{22}$ fixed to
various values in turn.
The upper curve applies to $\Delta_1$ and the lower curve
of the same line type, to $\Delta_2$. While in all cases $\Delta_1$
increases with increasing $\lambda_{12}$, in one case (solid curve),
the lower gap decreases slightly. More importantly, the value of the upper
gap ratio increases above its BCS ratio 3.53 and can
reach 4.6 in renormalized BCS, a feature which comes from the
two-band nature of the system. Comparing the dotted curves
to the solid circles for $\Delta_1$, we note that Eliashberg results are always
above their $\lambda^{\theta\theta}$ counterpart, reflecting well-known
strong coupling corrections to the gap. This applies as well to $\Delta_2$,
the lower gap. We now comment specifically on the other curves.
To increase the anisotropy between $\Delta_1$ and $\Delta_2$ for the 
parameter set considered here, we need to decrease the value of
$\lambda_{22}$. Note, however, that even when we assume a repulsion
in the second band, equal in size to the attraction $\lambda_{11}=1$ in 
the first band (long-dashed curve), a substantial gap is nevertheless
induced in the second channel even for $\lambda_{12}=0$. It is the
finite value of $\lambda_{21}$ which produces this gap. Recall that 
$\lambda_{21}$ describes the effect of band 1 on band 2
due to interband electron-phonon coupling. Turning on, as well, some
$\lambda_{12}$ increases the second gap further but not by much.
Finally, we mention that  as $\lambda_{21}$ increases (not
shown here),  $\Delta_1$ decreases while
 $\Delta_2$ increases, {\it ie.} the ratio  of $\Delta_2/\Delta_1$ goes up
towards one and the anisotropy is reduced.

\begin{figure}[ht]
\begin{picture}(250,200)
\leavevmode\centering\includegraphics{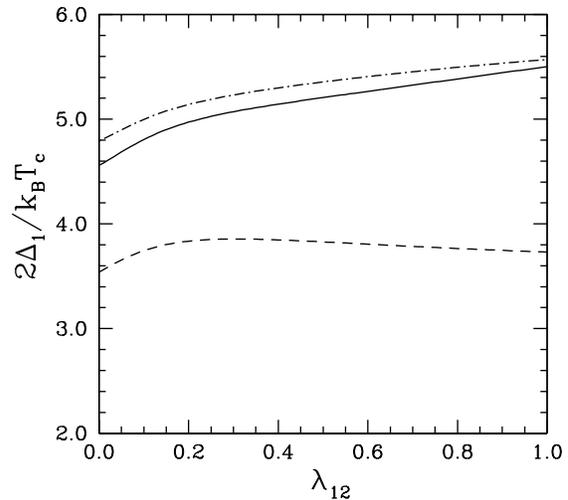}
\end{picture}
%\vskip 70pt
%\begin{figure}
%\includegraphics[clip,width=0.45\textwidth]{tc1.eps}
%\includegraphics{tc1.eps}
\caption{Gap ratio $2\Delta_1/k_BT_c$
as a function of $\lambda_{12}=\lambda_{21}$,
for 
$\lambda_{11}=1.3$ and $\lambda_{22}=0.5$. These curves
provide a comparison between the Eliashberg calculation 
(solid curve)
and the RBCS calculation (dashed curve), 
along with  the result from using the RBCS
expression with the
strong coupling correction formula given in the text (dot-dashed curve).}
\label{fig4}
\end{figure}

In Fig.~\ref{fig3}, the ratio $\lambda_{12}/\lambda_{21}=N_2/N_1$ is
varying, while in Fig.~\ref{fig4}, we keep $\lambda_{12}=\lambda_{21}$ and
illustrate more clearly the effect of strong coupling Eliashberg
in comparison with the RBCS calculation, and also provide a comparison
with the RBCS calculation corrected with the strong coupling
formula of Eq.~(\ref{eq:etaDel}). One finds that the 
gap in Eliashberg is quite enhanced over the RBCS result, even
exhibiting a different qualitative behaviour with the Eliashberg
gap (solid curve) increasing with increasing offdiagonal $\lambda$
while the RBCS counterpart (dashed curve)
is decreasing. However, when the strong
coupling correction formula is applied to the RBCS result, the resulting
curve (dot-dashed) is now in reasonable
agreement with the Eliashberg calculation and
follows the evolution with increasing offdiagonal $\lambda$ very well.

It is of interest to experimentalists\cite{wei}, looking at novel materials
suspected of harbouring multiband superconductivity, whether
there may be a range of parameters that could produce a very large upper
gap ratio with a large anisotropy in magnitude between the upper and lower
gaps.
It is possible that it could occur in a regime where $\lambda_{12}/\lambda_{21}\gg 1$,
 as suggested by the trend in our Fig.~\ref{fig3},
while in the opposite regime we will show that all results return to
standard weak coupling BCS values. As previously mentioned, this ratio of $\lambda_{12}/\lambda_{21}$
is equivalent to the ratio of density of states in the two bands,
sometimes denoted as $\alpha$ in the literature,
{\it i.e.} $\alpha\equiv\lambda_{12}/\lambda_{21}=N_2/N_1$.
We have gone to $\alpha=20$ within the renormalized BCS formalism and
were not able to
 produce gap ratios bigger than about 5 or so, for the parameters
examined,
and at the same time, the lower gap ratio was about
3. We conclude, therefore, that even with added strong
coupling effects, very large gap ratios
tending towards 10 to 20 are difficult to obtain in
conjunction with a large anisotropy in the two gaps. Repulsive potentials in
the second band can give a large anisotropy, but they also lower the value of
the upper gap ratio. 
Later in Section~VI, we will return to this issue
of trying to obtain large gap ratios and large gap anisotropy, when we
examine another extreme limit first considered by Suhl {\it et al.}\cite{suhl}.

To conclude this subsection, we examine an approximate formula for the
gap ratio in two-band superconductivity, which has been given and
used by experimentalists\cite{iavarone}, 
to determine its range of validity in the face of more
exact calculations. The formula is an unrenormalized BCS formula
and we have already seen that renormalization and strong
coupling effects can be substantial. 
For $\lambda_{22},\lambda_{12},\lambda_{21}\ll\lambda_{11}$, we can
derive the primary (or large) gap ratio as:
\begin{eqnarray}
\frac{2\Delta_1}{k_BT_c}&\simeq& 3.53\biggl[1-\frac{\lambda_{12}}{\lambda_{21}}
u^2\ln u\biggr]\nonumber\\
&=&3.53\biggl[1-\frac{N_{2}}{N_{1}}
\biggl(\frac{\Delta_2}{\Delta_1}\biggr)^2\ln \biggl(\frac{\Delta_2}{\Delta_1}\biggr)\biggr],
\label{eq:ivar}
\end{eqnarray}
which is the same equation as given in Iavarone {\it et al.}\cite{iavarone},
where their use of the indices 1 and 2 are reversed with respect to ours.
In our formula (\ref{eq:ivar}) given here, the $u$ and $\lambda$'s
are coupled through Eq.~(\ref{eq:u}), but in the case of Ref.\cite{iavarone}
the ratio of the density of states and the ratio of the gaps are
treated as independent parameters with the only constraint being
that $u\ll 1$.

\begin{figure}[ht]
\begin{picture}(250,200)
\leavevmode\centering\includegraphics{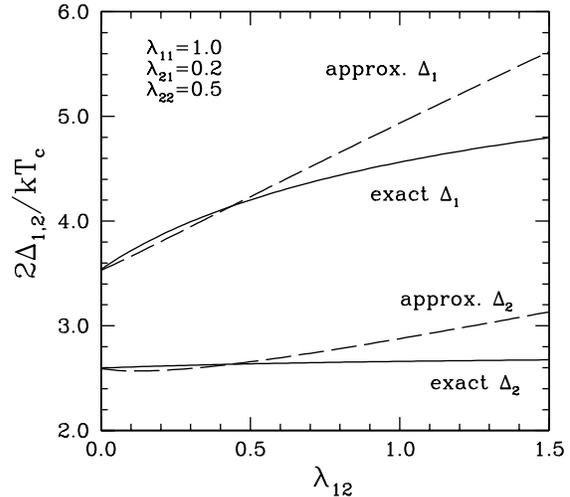}
\end{picture}
%\vskip 60pt
%\begin{figure}
%\includegraphics[clip,width=0.45\textwidth]{tc1.eps}
%\includegraphics{tc1.eps}
\caption{Gap ratios for the upper ($2\Delta_1/k_BT_c$)
and lower gap ($2\Delta_2/k_BT_c$) as a function of $\lambda_{12}$
for $\lambda_{11}=1.0$, $\lambda_{22}=0.5$,
 $\lambda_{21}=0.2$. The solid curve is the exact BCS result,
whereas, the dashed curve illustrates the approximate formula of
Iavarone {\it et al.}\cite{iavarone}.}
\label{fig5}
\end{figure}

In Fig.~\ref{fig5}, we compare this approximate BCS formula with
that of our exact renormalized BCS formula for typical $\lambda_{ij}$
values used in the literature. The $\mu^*_{ij}$ are set to zero as
there is no such feature in the Iavarone {\it et al.}
formula and the $\mu^*$'s in that case would simply serve to 
change the effective value of $\lambda$'s. We find that the approximate
formula (dashed curve of Fig~\ref{fig5})
 compares well with the renormalized BCS result in
the limit of small $\lambda_{12,21,22}$, as required by the constraint
of the approximation, and breaks down for $\lambda_{12}>0.5$, where
the approximate formula tends to overestimate quite significantly the
value of the two gaps. Strong coupling effects would produce very significant
deviations in addition. Not shown is the case where $\lambda_{12,21,22}$
were all taken to be very small and then in that case, as expected, there
was excellent agreement between the exact renormalized BCS calculation
and the approximate form. The fact that Iavarone {\it et al.}\cite{iavarone}
obtained excellent estimates of the two energy
gaps for MgB$_2$ is maybe fortuitous in 
some sense, because it will be seen in the next section, where we discuss
MgB$_2$ in detail, that the renormalized BCS  formula underestimates
the correct gap values of MgB$_2$ and strong coupling corrections of about 
7-10\%
are needed to obtain good agreement between the data and full Eliashberg calculations. We conclude that their simple formula is helpful,
but that it should be used with caution when
considering systems where the parameters are no longer small as then this
formula will fail. 

\subsection{Specific Heat Jump}

The specific heat is calculated from the free energy. The
difference in free energy $\Delta F=F_S-F_N$ between the superconducting state and
the normal state is given by\cite{carbotte}:
\begin{eqnarray}
\Delta F &=& -\pi T\sum_{n=-\infty}^{+\infty}\sum_i N_i(0)\bigl[\sqrt{\omega_n^2+\Delta^2_i(i\omega_n)}
-|\omega_n|\bigr]\nonumber\\
&\times& \biggl[Z^S_i(i\omega_n)-Z^N_i(i\omega_n)\frac{|\omega_n|}{\sqrt{\omega_n^2+\Delta^2_i(i\omega_n)}}\biggr],
\label{eq:freeenergy}
\end{eqnarray}
where ``S'' and ``N'' refer to the superconducting and normal state, 
respectively, and $i$ indexes the number of bands. From this, the
difference in the specific heat is obtained:
\begin{equation}
\Delta C=-T\frac{d^2\Delta F}{dT^2},
\end{equation}
and the negative of the slope of the difference in specific heat near $T_c$ is given as
\begin{equation}
g = -\frac{d\Delta C(T)}{dT}\bigg|_{T_c}\frac{1}{\gamma},
\end{equation}
where $\gamma$ is the Sommerfeld constant for the two-band case.

In the $\lambda^{\theta\theta}$ model,
the gap near $T_c$, for $t=T/T_c$, can be written as
\begin{eqnarray}
\Delta_1^2(t)=\frac{8(\pi T_c)^2}{7\zeta(3)}\frac{\eta_C}{\chi_1}(1-t),\label{eq:gapt}\\
\Delta_2^2(t)=\frac{8(\pi T_c)^2}{7\zeta(3)}\frac{1}{\chi_2}(1-t),
\end{eqnarray}
where $\zeta(3)\simeq 1.202$.
Here,
\begin{equation}
\chi_1=\frac{(1-A\bar\lambda_{22})\bar\lambda_{11}
+\bar\lambda_{12}\bar\lambda_{21}A[1+A^2\bar\lambda_{21}^2(1-A\bar\lambda_{22})^{-3}]}
{(1-A\bar\lambda_{22})\bar\lambda_{11}+\bar\lambda_{12}\bar\lambda_{21}
[2A+A^2\bar\lambda_{22}/(1-A\bar\lambda_{22})]}
\end{equation}
and
\begin{equation}
\chi_2=\frac{(1-A\bar\lambda_{11})\bar\lambda_{22}
+\bar\lambda_{21}\bar\lambda_{12}A[1+A^2\bar\lambda_{12}^2(1-A\bar\lambda_{11})^{-3}]}
{(1-A\bar\lambda_{11})\bar\lambda_{22}+2A\bar\lambda_{21}\bar\lambda_{12}
[2A+A^2\bar\lambda_{11}/(1-A\bar\lambda_{11})]},
\end{equation}
and the strong coupling correction is introduced through\cite{marcar}:
\begin{equation}
\eta_C = 1+53\biggl(\frac{T_c}{\omega_{\ln}}\biggr)^2\ln\biggl(\frac{\omega_{\ln}}
{3T_c}\biggr).
\end{equation}
The specific heat jump at $T_c$ is:
\begin{equation}
\frac{\Delta C}{\gamma T_c} =1.43\Biggl[
\frac{(1+\lambda_{11}+\lambda_{12})\frac{\eta_C}{\chi_1}+\alpha(1+\lambda_{22}+\lambda_{21})\frac{1}{\chi_2}}
{(1+\lambda_{11}+\lambda_{12})+\alpha(1+\lambda_{22}+\lambda_{21})}\Biggr].
\label{eq:specfull}
\end{equation}
We find with this expression
that anisotropy (ie., $\lambda_{11}\ne\lambda_{22}$) reduces the jump
ratio but
increasing $\lambda_{12}$
or  $\lambda_{21}$ 
increases the ratio, and the maximum
obtainable is 1.43. 
Other work along the same line is given in Refs.~\cite{kita,mishonov}
where they do not consider full renormalized BCS or strong coupling
theories, as we have done here.

When $\lambda_{12},\lambda_{21}\to 0$, $1/\chi_1\sim 1+O(\bar\lambda^2_{12})$
and $1/\chi_2\sim O(\bar\lambda^4_{12})$. This is assuming
$\bar\lambda_{11}-\bar\lambda_{22}$ and
$\bar\lambda_{22}$ remain significant as compared with the
value of the offdiagonal elements. In this case,
\begin{equation}
\frac{\Delta C}{\gamma T_c} =1.43\biggl[
\frac{(1+\lambda_{11}+\lambda_{12})}
{(1+\lambda_{11}+\lambda_{12})+\alpha(1+\lambda_{22}+\lambda_{21})}\biggr].
\label{eq:specsimple}
\end{equation}
The physics of this formula is that, in this limit, the specific heat
jump at $T_c$ itself is determined only by the superconductivity of the
dominant band, but it is normalized with the normal state specific heat
$\gamma$ belonging to the sum of both bands. This has the effect of making
$\Delta C(T_c)/\gamma T_c$ always less than the BCS value by a factor
of $1/(1+\alpha^*)$, where
$\alpha^*=\alpha(1+\lambda_{22}+\lambda_{21})/(1+\lambda_{11}+\lambda_{12})$. 
For MgB$_2$, we expect $\alpha^*\buildrel >\over\sim 1$ which means that
in this case the normalized jump is reduced to about half its BCS value.
If we had included in (\ref{eq:specsimple}) the strong coupling correction
$\eta_C$, 
this would have the effect of increasing the factor 1.43 to a larger value
characteristic of strong coupling but the additional anisotropy parameters
would still work to reduce the jump. Thus, in a two-band superconductor,
the jump will be smaller than for one band with the same strong coupling index 

\subsection{Thermodynamic Critical Magnetic Field}

The thermodynamic critical magnetic field is calculated from the
free energy difference:
\begin{equation}
H_c(T)=\sqrt{-8\pi\Delta F}.
\end{equation}
As the temperature dependence of this quantity, normalized to its
zero temperature value, follows very closely a nearly quadratic behaviour,
the deviation function $D(t)$ is often plotted:
\begin{equation}
D(t)\equiv \frac{H_c(T)}{H_c(0)}-(1-t^2),
\label{eq:devfn}
\end{equation}
where $t=T/T_c$.

At $T=0$
\begin{equation}
H_c^2(0)=4\pi N_1^*\Delta_1^2(1+\alpha^*u^2),
\label{eq:Hc0}
\end{equation}
where $\alpha^*=N_2^*/N_1^*$ and
\begin{equation}
N_i^*=N_i(0)(1+\lambda_{ii}+\lambda_{ij}).
\end{equation}
The zero temperature critical magnetic field is
modified through the second term in (\ref{eq:Hc0}) which increases with 
increasing $\alpha^*$ and with the 
square of the anisotropy ratio $u$, which in this case is just the
ratio of the independent gap values for the two separate bands.
Further, the dimensionless ratio is
\begin{equation}
\frac{\gamma T_c^2}{H_c^2(0)}=\frac{\pi(k_BT_c)^2[1+\alpha^*]}
{6\Delta_1^2[1+\alpha^*u^2]}.
\label{eq:hratio}
\end{equation}
For almost decoupled bands, Eq.~(\ref{eq:hratio}) becomes
\begin{equation}
\frac{\gamma T_c^2}{H_c^2(0)}=0.168\frac{1+\alpha^*}
{1+\alpha^*u^2},
\label{eq:hratio2}
\end{equation}
where the second factor on the right-hand side modifies the usual
single-band BCS value of
0.168 for the presence of the second band. Again, both $\alpha^*$ and
$u$ enter the correction.
If there is no anisotropy, $u=1$, and therefore the bands must be the same,
we recover the one-band limiting value. For
large anisotropy where $u\to 0$, and if $\alpha^*$ is of order one,
the ratio in Eq.~(\ref{eq:hratio2}) is of order twice its one-band value
because the second band contributes very little to the zero temperature
condensation energy, but is still as equally important as the first band
in its contribution to $\gamma T_c$, the normal state specific heat. 
Near $T_c$
\begin{equation}
H_c(t)=\sqrt{\frac{32\pi}{7\zeta(3)}}(\pi k_BT_c)(1-t)\biggl[
\frac{N_1^*}{\chi^2_1}+\frac{N_2^*}{\chi_2^2}\biggr]^{1/2},
\label{eq:Hct}
\end{equation}
which then gives the dimensionless ratio
\begin{eqnarray}
h_c(0)&\equiv& \frac{H_c(0)}{|H_c'(T_c)|T_c}\nonumber\\
&=& \frac{2\Delta_1}{k_BT_c}\frac{1}{\pi}\sqrt{\frac{7\zeta(3)}{32}}
\sqrt{\frac{1+\alpha^*u^2}{\chi_1^{-2}+\alpha^*\chi_2^{-2}}}.
\label{eq:hc0}
\end{eqnarray}
Strong coupling factors could be introduced in (\ref{eq:Hc0}),
(\ref{eq:hratio}), and (\ref{eq:Hct}). They are not
given explicitly here as they are less important than for the specific 
heat jump and the slope of the penetration depth at $T_c$ (see Table~\ref{table1}).
The limit of nearly decoupled bands 
($\bar\lambda_{12},\bar\lambda_{21}\ll\bar\lambda_{11}
-\bar\lambda_{22},\bar\lambda_{22}$)
gives for this quantity:
\begin{equation}
h_c(0)=0.576\sqrt{1+\alpha^*u^2}.
\label{eq:odhc0}
\end{equation}
The square root, which accounts for two-band effects contains a correction
proportional to $\alpha^*u^2$. It can be understood as follows.
The slope at $T_c$ found from formula (\ref{eq:Hct})
depends only on band 1 but $H_c(0)$ involves  both
and hence this correction comes solely from
$H_c(0)$ as seen in Eq.~(\ref{eq:Hc0}). If the anisotropy
between the two bands is large $u\to 0$, there is no correction factor in
(\ref{eq:odhc0}) because  the second band is eliminated from $H_c(0)$.
If, on the other hand, $u$ is near 1, the two bands have nearly equal
gap value but still it is only band 1 which contributes to the slope
at $T_c$ and the dimensionless ratio (\ref{eq:odhc0}) can now 
be larger than its
BCS value. 

\subsection{Penetration Depth}

The London penetration depth $\lambda_L(T)$ is evaluated from\cite{carbotte}:
\begin{equation}
\frac{1}{\lambda_L^2(T)}=\frac{T}{2}\sum_{n=1}^{\infty}\sum_i\frac{1}{\lambda^2_{ooi}}
\frac{\Delta^2_i(i\omega_n)}{Z_i(i\omega_n)[\omega_n^2+\Delta^2_i(i\omega_n)]^{3/2}},
\label{eq:pendepth}
\end{equation}
where in three dimensions
\begin{equation}
\frac{1}{\lambda_{ooi}^2}=\frac{4\pi n_ie^2}{m_ic^2}=\frac{8\pi e^2}{3c^2}N_iv_{Fi}^2
\end{equation}
and $v_{Fi}$ is the Fermi velocity in the band labelled by the index $i$.
This last equation would be multiplied by a factor of 3/2 in two dimensions.

For the penetration depth $\lambda_L(T)$ at $T=0$,
\begin{equation}
\frac{1}{\lambda_L^2(0)}=\frac{1}{\eta^2_{\lambda_L}(0)\lambda_{oo1}^2(1+\lambda_{11}+\lambda_{12})}+\frac{1}{\lambda_{oo2}^2(1+\lambda_{22}+\lambda_{21})},
\end{equation}
and near $T_c$,
\begin{eqnarray}
\frac{1}{\lambda_L^2(t)}&=&2(1-t)\biggl[\frac{1}{\eta^2_{\lambda_L}(T_c)\lambda_{oo1}^2\chi_1(1+\lambda_{11}+\lambda_{12})}\nonumber\\
&+&\frac{1}{\lambda_{oo2}^2\chi_2(1+\lambda_{22}+\lambda_{21})}\biggr],
\end{eqnarray}
where
\begin{eqnarray}
\eta_{\lambda_L}(0)&=&1+1.3\Biggl(\frac{T_c}{\omega_{\ln}}\Biggr)^2\ln\Biggl(
\frac{\omega_{\ln}}{13T_c}\Biggr),\\
\eta_{\lambda_L}(T_c)&=&1-16\Biggl(\frac{T_c}{\omega_{\ln}}\Biggr)^2\ln\Biggl(
\frac{\omega_{\ln}}{3.5T_c}\Biggr).
\end{eqnarray}

Hence, defining $y_L(T)=1/\lambda_L^2(T)$, we write the 
dimensionless BCS penetration
depth ratio $y$ as
\begin{equation}
y \equiv \frac{y_L(0)}{|y_L'(T_c)|T_c}=\frac{1}{2}(1+\alpha\beta)\biggl[\frac{1}{\chi_1}
+\frac{\alpha\beta}{\chi_2}\biggr]^{-1},
\label{eq:yratio}
\end{equation}
where $\beta=v^2_{F2}(1+\lambda_{11}+\lambda_{12})/v^2_{F1}(1+\lambda_{22}+\lambda_{21})$.
$\beta$ is expected to be of order 1 unless there is a great disparity in
the two Fermi velocities. For MgB$_2$, we use the values of $v_{F1}=4.40\times
10^5$ m/s and $v_{F2}=5.35\times 10^5$ m/s
reported in Ref.~\cite{golubov} and for our other model calculations,
we take them to be equivalent, for simplicity.
For the nearly decoupled case
\begin{equation}
\frac{y_L(0)}{|y_L'(T_c)|T_c}=\frac{1}{2}(1+\alpha\beta).
\end{equation}
For $\alpha$ and $\beta$ equal to one, we see that the normalized slope
of the penetration depth is twice its one-band BCS value of $1/2$. Should $\alpha$,
$\beta$, or both be much larger than 1, then the slope can be even larger, 
which reflects the fact that only the dominant band determines the slope $y_L'$
but both bands contribute to $y_L(0)$. Information on the $v_{Fi}$ and the
$N_i(0)$ is
contained in the slope.

\begin{figure}[ht]
\begin{picture}(250,200)
\leavevmode\centering\includegraphics{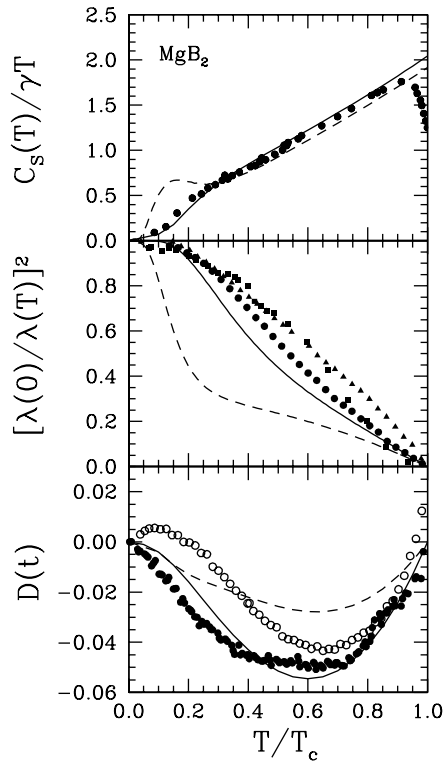}
\end{picture}
\vskip 70pt
%\begin{figure}
%\includegraphics[clip,width=0.45\textwidth]{tc1.eps}
%\includegraphics{tc1.eps}
\caption{Upper frame: Electronic specific heat for MgB$_2$
in the superconducting state normalized to the normal state
as a function of $T/T_c$. The points are the experimental results
of Wang {\it et al.}\cite{junod} and the solid curve is the result for
the Eliashberg calculation using the parameters given in the 
literature\cite{mitrovic}. The dashed curve illustrates the case where the
$\lambda_{12}$ and $\lambda_{21}$ parameters, used for the solid curve,
have been halved. The jump due to the lower gap begins to appear in this
case. Middle frame: $[\lambda(0)/\lambda(T)]^2$ versus $T/T_c$. Curves
are those resulting from the same set of parameters as discussed for the
upper frame, with the $v_{Fi}$
taken from Ref.~\cite{golubov}. 
The data, shown for comparison, have been taken from
Ref.~\cite{dolgovpen}. No impurity scattering has been used to obtain
a better fit.
Lower frame: The deviation function $D(t)$ for the thermodynamic critical
field. Line labels are as above and the data (open and solid circles) 
are formed
from the $H_c(T)$ data given by Wang {\it et al.}\cite{junod}
and Bouquet {\it et al.}\cite{bouquet}, respectively.}
\label{fig6}
\end{figure}

\section{M\lowercase{g}B$_2$: integrated bands and strong coupling}

We  now continue on beyond renormalized BCS formulas to evaluate
quantities based on the full two-band Eliashberg formalism and
we begin with the specific 
case of MgB$_2$
and strong coupling effects. Eqs.~(\ref{eq:Del}) and (\ref{eq:Z})
were solved for electron-phonon spectral densities $\alpha^2_{ij}F(\omega)$,
read from graphs in Ref.~\cite{mitrovic}, which were originally
presented in Ref.~\cite{goliop}. The Coulomb
repulsion parameters $\mu^*_{ij}$ and $\lambda_{ij}$,
taken from \cite{goliop},
were: $\lambda_{\sigma\sigma}=1.017$,
$\lambda_{\pi\pi}=0.448$, $\lambda_{\sigma\pi}=0.213$, 
$\lambda_{\pi\sigma}=0.155$, 
$\mu^*_{\sigma\sigma}=0.210$, $\mu^*_{\pi\pi}=0.172$, 
$\mu^*_{\sigma\pi}=0.095$, and $\mu^*_{\pi\sigma}=0.069$, with
$\omega_c=750$ meV. From these parameters, $T_c$ was found to be 39.5~K.
As discussed in our theory introduction,
we used $\omega_{\ln}=66.4$ meV, calculated from the
$\alpha^2_{11}F(\omega)$ spectrum, to form
our strong coupling index $T_c/\omega_{\ln}$.
The other three channels had $\omega_{\ln}\simeq 62$ meV,
which is not so different, although as argued previously, the
main strong coupling effects will come from the 11 channel, and
hence the choice of 66.4 meV for this parameter.
From the solution of the Eliashberg equations, 
we can evaluate Eq.~(\ref{eq:freeenergy}) 
for the free energy difference between the
superconducting and normal state, and evaluate the superfluid density
or the 
inverse square of the penetration depth from Eq.~(\ref{eq:pendepth}). 
In Fig.~\ref{fig6}, which has three frames: the top is the specific
heat, middle, the penetration depth, and bottom, the critical magnetic field
deviation function of formula (\ref{eq:devfn}), we compare Eliashberg results
(solid curve) with experimental results 
(solid and open circles, triangles, and squares).

In all cases, the agreement with experiment is very good and certainly
as good as is obtained in conventional one-band cases\cite{carbotte}.
In each case, we also present a second set of theoretical results
(dashed curve) for which all microscopic parameters remain those of MgB$_2$
except that we have half the value of the offdiagonal spectral functions
$\alpha^2_{12}F(\omega)$ and $\alpha^2_{21}F(\omega)$, which changes
the $T_c$ only by about one degree. 
It is clear that doing this
reduces greatly the quality of the fit one obtains with the experimental
data. This can be taken as evidence that the electronic  structure,
first-principle calculations of electron-phonon spectral functions are 
accurate. It also shows that variation of parameters by a 
factor of two or so away from the computed ones can lead to significant  
changes
 in superconducting properties and, in this instance,
features of the second transition,
due to the lower gap, begin to appear. 
The specific heat curve was computed before
in Refs.~\cite{choi,choinature} and the penetration depth in Refs.~\cite{dolgovpen,moca}. In these cases, our calculations
(solid curves) confirm previous ones 
and demonstrate that our calculational procedure is
working correctly. For the penetration depth
we did not introduce impurity scattering. Impurities can affect the
penetration depth and were included in Ref.~\cite{dolgovpen}. The three
sets of penetration depth 
data are for clean (solid circles\cite{jin} and triangles\cite{manzano})
and dirty samples (solid squares\cite{nie})  as discussed in \cite{dolgovpen}.
To our knowledge, the deviation function has not been computed and compared
with experiment before and it is presented for the first time
here. The data is from Refs.~\cite{junod} (open circles) and \cite{bouquet}
(solid circles) and again agreement with calculation, with no free
parameters, is very good. The minimum in the deviation function
for the Eliashberg calculation occurs at $T/T_c=0.6$ and has a value
of -0.054. In the experimental data, the minima occur at about $T/T_c=0.6$
and 0.65, with values of about -0.05 and -0.045, respectively. For
reference, the one-band BCS value is -0.037 and
strong coupling makes this value even smaller and can even
push it to a positive value, hence anisotropy is 
compensating for the strong coupling effects and is making this value
larger and more negative.\cite{carbotte}

\begin{figure}[ht]
\begin{picture}(250,200)
\leavevmode\centering\includegraphics{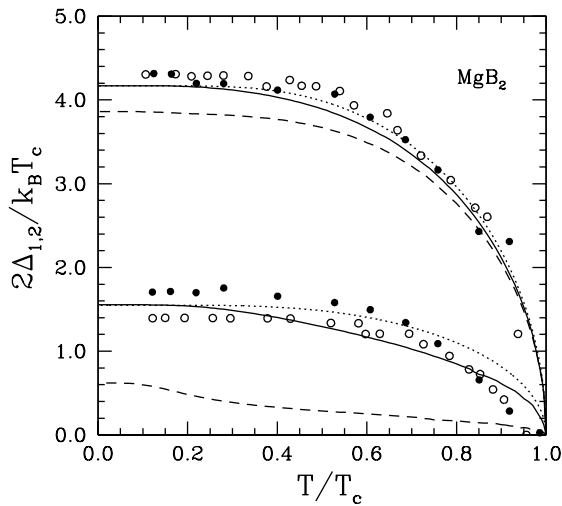}
\end{picture}
%\vskip 60pt
%\begin{figure}
%\includegraphics[clip,width=0.45\textwidth]{tc1.eps}
%\includegraphics{tc1.eps}
\caption{Gap ratios for the upper ($2\Delta_1/k_BT_c$)
and lower gap ($2\Delta_2/k_BT_c$) as a function of $T/T_c$.
Shown as the solid curves are the predictions for the gap
ratios given by our full Eliashberg calculations for MgB$_2$,
the dashed curves are the Eliashberg calculations for the case
of reducing the offdiagonal $\lambda$'s by half and the dotted
curves show the classic BCS temperature dependences to illustrate
the deviation of the temperature dependence of the Eliashberg
two-band calculation for MgB$_2$. The open circles are the data
from
Iavarone {\it et al.}\cite{iavarone}, where we have used a $T_c=38.3K$
to obtain their quoted upper gap ratio value of 4.3. The solid dots
are the data of Gonnelli {\it et al.}\cite{gonnelli}.}
\label{fig7}
\end{figure}

In Fig.~\ref{fig7}, we present the temperature dependence of the
two gap ratios for MgB$_2$. Once again the solid curve is the
full Eliashberg calculation using the parameters given for MgB$_2$
with no adjustments. 
The ratio $\Delta_1/\Delta_2$ increases from 2.7 at $T=0$ to about 3.5
at $T_c$.
The temperature-dependent behaviour shown here was also found by
Choi {\it et al.}\cite{choinature},
Brinkman {\it et al.}\cite{golubov},
and Golubov {\it et al.}\cite{goliop}.
A comparison with some of the more recent experiments
is given by the open and closed circles, 
with the data taken from Iavarone {\it et al.}\cite{iavarone}
and Gonnelli {\it et al.}\cite{gonnelli}, respectively.
Similar data is found in other references\cite{szabo,monod,schmidt}. 
In the case
of the data by Iavarone {\it et al.}, the statement of $T_c$ was ambiguous
and so we used their quoted value of the upper gap ratio of 4.3 along
with their quoted value of the upper gap being 7.1 meV to determine a 
$T_c=38.3K$ used for the scaling of the data for the plot presented here.
The Gonnelli {\it et al.} data is presented based on the $T_c$ of 38.2K
given in their paper. There is a very reasonable agreement of the data
with the calculation, once again, along with Fig.~\ref{fig6}, this shows
a consistency of a number of sets of data from several different
types of experiments with the one set of parameters 
fixed from band structure for MgB$_2$. Thus overall, the agreement between
theory and experiment is excellent and validates the two-band nature
of superconductivity in this material. The dotted curves in Fig.~\ref{fig7} are
presented to show that the two-band calculations do show deviation from
a classic BCS temperature dependence (which was used in the original
presentations of the data\cite{iavarone,gonnelli}). In particular,
Gonnelli {\it et al.} argued that the deviation of their lower gap data
at temperatures above 25K (or $T/T_c=0.65$, here) from the BCS temperature
dependence is an additional signature of the two-band nature of the
material. However, we find no such dramatic suppression in the two-band
calculations at this temperature and only with the dashed curve, where
we have taken the offdiagonal electron-phonon coupling to be half of
the usual value for MgB$_2$ do we find an inflection point around
 0.35. We were not able to induce a suppression of the lower
gap in the vicinity of $T_c$ by varying the MgB$_2$ parameters
slightly about their accepted values. However, such behaviour can be
found in other regimes of the parameter space not relevant to MgB$_2$
and this feature and the issue raised by Gonnelli {\it et al.} will be 
discussed further in the next section. To end, note that
 an inflection point is also 
seen in the penetration depth at
about $T/T_c\sim 0.35$, as described first by Golubov {\it et al.}\cite{dolgovpen}
and also found here (solid curve of middle frame of Fig.~\ref{fig6}).

\begin{table*}
\caption{Universal dimensionless BCS ratios and their modification for strong coupling (SC)
and two-band superconductivity. RBCS stands for Renormalized BCS formula
given in text. The percentage difference between the full Eliashberg (Eliash.)
calculation and RBCS, used to measure the amount of strong coupling
correction, is given as \% SC and defined as $|({\rm Eliash.}-{\rm RBCS})/
{\rm Eliash.}|$. }
\begin{ruledtabular}
\begin{tabular}{cccccccccccc} % In second brace, l = left, r = right,
% c = centered and d = decimal justification.
Ratio & BCS & Pb & MgB$_2$ & MgB$_2$ & MgB$_2$ & MgB$_2$ & MgB$_2$ & Lor &Lor &Lor &Lor\\  
 &one-band & one-band & Eliash.& Expt. & RBCS & \% SC & RBCS$+$SC & Eliash.&RBCS & \% SC &RBCS$+$SC\\  
% Separate items with &. End line with \\
 \hline % Creates a horizontal line.
 $T_c/\omega_{\ln}$ & 0.0 & 0.128 & 0.051& 0.076\footnotemark[1] & 0.0 &  & 0.052 &0.15&0.0& &0.15\\
 $2\Delta_1/k_BT_c$ & 3.53 & 4.49 & 4.17 & 3.6-4.6\footnotemark[2]& 3.86 & 7.4\%& 4.15&4.97&3.84&23\%&5.14\\
 $2\Delta_2/k_BT_c$ & 3.53 & 4.49 & 1.55 & 1.0-1.9\footnotemark[2]& 1.40 & 9.7\%&&2.66&2.27&15\%&\\
 $\Delta_2/\Delta_1$ & 1.00 & 1.00 & 0.37 &  0.30-0.42\footnotemark[2]& 0.36&2.7\% & &0.535&0.593&11\%& \\
$\Delta C/\gamma T_c$ & 1.43 & 2.79 & 1.04 &  0.82-1.32\footnotemark[3]& 0.817 &21\% & 1.02 &2.08 &1.07 &49\% &1.97\\
$g$ & -3.77 & -12.68 & -3.28 & -(2.37-4.31)\footnotemark[4] & & &  &-8.32&  & & \\
$\gamma T_c^2/H_c^2(0)$ & 0.168 & 0.132 & 0.225 & 0.183\footnotemark[5]& 0.247 & 9.8\%& & 0.153&0.193&26\% & \\
$h_c(0)$ & 0.576 & 0.465 & 0.581 & 0.518-0.667\footnotemark[6]& 0.629 & 8.3\%& & 0.500&0.621&24\% & \\
$y$ & 0.5 & 0.311 & 1.25 &1.22\footnotemark[7],0.547\footnotemark[8] & 1.50 & 20\%&1.32 & 0.536&0.861&61\% & 0.569\\
\end{tabular}
\label{table1}
\end{ruledtabular}
\footnotetext[1]{Ref.~\cite{junod}}%Tcwln
\footnotetext[2]{Refs.~\cite{bouqueteuro,putti,iavarone,gonnelli,szabo,monod,schmidt}}%gap1,gap2,u
\footnotetext[3]{Refs.~\cite{bouquet,putti,junod,yang}}%spec jump
\footnotetext[4]{Estimated from Refs.~\cite{junod,bouquet,putti}}%g
\footnotetext[5]{Ref.~\cite{bouquet}}%hratio
\footnotetext[6]{Estimated from Refs.~\cite{junod,bouquet,kang}}%hc0
\footnotetext[7]{Estimated from data of Ref.~\cite{jin} as presented in \cite{dolgovpen}}%y
\footnotetext[8]{Estimated from Ref.~\cite{manzano}}%y
\end{table*}

More results from our calculations as well as comparison with data are
presented in Table~\ref{table1}. In the first column,
we include, for comparison, the one-band BCS values for the various dimensionless
ratios. The strong coupling index is first, followed by the major gap to
critical temperature ratio, the minor gap ratio, 
the anisotropy $\Delta_2/\Delta_1$,
the normalized specific heat jump and the negative of its slope at $T_c$,
$\gamma T_c^2/H^2_c(0)$, and the inverse of the normalized
slope at $T_c$ for the critical magnetic field and for the penetration
depth. Included in the second column, also for comparison, are the same indices
for Pb, the prototype, single-band, strong coupler. 
We remind the reader that, in
many conventional superconductors, 
strong coupling corrections are large and that the data
cannot be understood without introducing them,
and these are to be differentiated from those corrections due to
anisotropy. The third column gives the 
results of our two-band calculations for MgB$_2$. This is followed by a column
giving experimental values. It is clear that the agreement between theory and
experiment is good. Note that we have not attempted to make
a complete survey of all experiments, but have tried to
present as many as reasonable, with no judgement
about the quality of the data or samples, which might
have improved over time. In addition,
for the quantities related to slopes, {\it i.e.}, $g$, $h_c(0)$, and $y$,
 we have
tried to estimate these ourselves from the graphs in papers and so
this should be viewed as rough estimates as the values
might change with a more rigorous analysis of
the original data. Also shown are the results when our renormalized BCS
formulas of the previous section are implemented using
MgB$_2$ parameters\cite{mustar}, which allows us
to define a measure of 
strong coupling corrections, entered in column 6 as percentages.
It is seen that MgB$_2$ is an intermediate coupling case. 
The next column shows the results when the analytical expressions for strong
coupling corrections to renormalized BCS, given in the text, are
applied. This improves the agreement with the full Eliashberg results
as compared to RBCS. Some discrepancies remain due in part to additional
modifications introduced by the coupling of a strong coupling
band with  a weak coupling
one  through the offdiagonal $\lambda_{ij}$'s.
The next four
columns were obtained for our
Lorentzian spectral density model
with
$\lambda_{11}=1.3$, $\lambda_{22}=0.5$, $\lambda_{12}=\lambda_{21}=0.2$,
and $\mu^*_{ij}=0$. This was devised to have
a strong coupling index $T_c/\omega_{\ln}\sim 0.15$ which is slightly larger
than Pb and well within the range of realistic values for electron-phonon
superconductors. It is clear that strong
coupling corrections are now even more significant and cannot be ignored in a 
complete theory.

\begin{figure}[ht]
\begin{picture}(250,200)
\leavevmode\centering\includegraphics{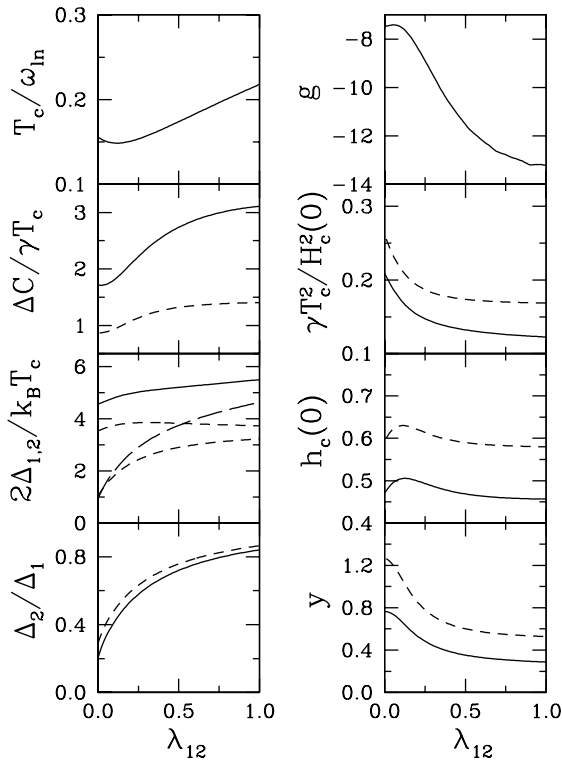}
\end{picture}
\vskip 70pt
%\begin{figure}
%\includegraphics[clip,width=0.45\textwidth]{tc1.eps}
%\includegraphics{tc1.eps}
\caption{Various BCS ratios as discussed in the text, shown as a
function of $\lambda_{12}$, where $\lambda_{21}=\lambda_{12}$
({\it i.e.} $\alpha=1$),
$\lambda_{11}=1.3$, and $\lambda_{22}=0.5$. The solid curves are
those for the full Eliashberg calculation for a Lorentzian model of
$\alpha^2F_{ij}(\omega)$ spectra and the short-dashed curves are for
the renormalized BCS formulas developed from the $\lambda^{\theta\theta}$
model and given in the text. For the frame with the gap ratios, the
upper gap is given by the solid curve and the lower gap is given by the
long-dashed curve, the upper and lower short-dashed curves are for
the upper and lower gaps, respectively, in RBCS. The first frame gives
the effective $T_c/\omega_{\ln}$ for the Eliashberg spectrum based on
the definition given in the text.}
\label{fig8}
\end{figure}

More information on strong coupling effects as well as on two-band anisotropy
is given in Fig.~\ref{fig8}, where we show the same BCS ratios 
as considered in Table~\ref{table1}. In all eight frames,
we have used our model Lorentzian $\alpha^2_{ij}F(\omega)$ spectra.
The solid curves are results of full Eliashberg calculations as a function
of $\lambda_{12}=\lambda_{21}$, with $\lambda_{11}$ fixed at 1.3
and $\lambda_{22}$ at 0.5. The dashed curves are for comparison and are
based on our $\lambda^{\theta\theta}$ formulas, {\it i.e}, give renormalized
BCS results
without use of the strong coupling
correction formulas. 
They, of course, can differ very significantly
from one-band universal BCS values because of the two-band anisotropy.
We see that these effects can be large and that on comparison between
the solid and dashed curves, the strong coupling effects can also be significant. 
As $\lambda_{12}=\lambda_{21}$ is increased from zero, with 
$\lambda_{11}$ and $\lambda_{22}$ remaining fixed, the $T_c$ increases and this
leads to the increase in $T_c/\omega_{\ln}$ from about 0.15 at 
$\lambda_{12}=\lambda_{21}=0$
to over 0.2 at $\lambda_{12}=\lambda_{21}=1$. 
For all the indices considered here, we note that
their values at $T_c/\omega_{\ln}=0.2$
are close to the values that they would have in a one-band case\cite{carbotte},
and the remaining anisotropy in the $\lambda_{ij}$'s play
only a minor role. (Of course, this is
a qualitative statement since it is well known that the shape of 
$\alpha^2F(\omega)$ for fixed $T_c/\omega_{\ln}$ can also affect somewhat the
value of BCS ratios\cite{carbotte}.) 
This is expected since
in this case the fluctuation off the average of any $\lambda_{ij}$ is becoming
smaller.
For RBCS all ratios have returned 
to
the one-band case at $\lambda_{12}=\lambda_{21}=1$ except for $y$
which remains 6\% larger.
We now comment on select indices separately. The normalized 
specific heat jump at $T_c$ in the $\lambda^{\theta\theta}$ model
is given by formula (\ref{eq:specfull}) with $\eta_C=1$. For
$\lambda_{12}=\lambda_{21}$ small, $\chi_1^{-1}\simeq 1+O(\bar\lambda^2_{12})$
and
$\chi_2^{-1}\simeq 0+O(\bar\lambda^4_{12})$. These conditions mean that 
$\Delta C/\gamma T_c$ rises slightly as $\lambda_{12}=\lambda_{21}$
increases, and eventually reaches 1.43.
By contrast, the solid
curve includes, in addition, strong coupling effects which increase the
value of the jump ratio 
rather rapidly. For $2\Delta_{1,2}/k_BT_c$, the lower gaps
have the same value for $\lambda_{12}=\lambda_{21}=0$ as it is determined only 
by $\lambda_{22}$. This is not so for the upper gaps. The dashed curve takes on
its BCS value of 3.53, but the solid curve (an Eliashberg calculation) has strong coupling 
effects as described in Fig.~\ref{fig4}. 
(This means that $\Delta_2/\Delta_1$
is smaller for the solid curve as compared to the dashed one
in the lower left-hand frame.) As $\lambda_{12}=\lambda_{21}$ increases,
the long-dashed and lower short-dashed curves begin to deviate because the
former starts to acquire strong coupling corrections of its own through
the offdiagonal $\lambda$'s. While the solid curve also increases, the 
anisotropy between 1 and 2 decreases. The short-dashed curves show different
behaviour. The ratio $2\Delta_1/k_BT_c$ starts at 3.53, rises slightly
towards 4 before tending towards 3.53 again. Now, the anisotropy between
$\Delta_2$ and $\Delta_1$ decreases mainly because $\Delta_2$ itself
rises towards 3.53. The behaviour of $\gamma T_c^2/H^2_c(0)$ (dashed
curve) can be understood from Eq.~(\ref{eq:hratio}). While $\Delta_1/T_c$,
as we have seen, does change somewhat with $\lambda_{12}=\lambda_{21}$,
a more important  change is the $u^2$ factor in the denominator of
(\ref{eq:hratio}) which rapidly decreases this ratio towards its BCS value
of 0.168 as $u$ increases towards 1. 
The behaviour of $h_c(0)$ given by Eq.~(\ref{eq:hc0}) is
more complex. The numerator in the square root goes towards $1+\alpha^*$,
as $u^2\to 1$, more rapidly than the denominator which involves the $\chi$'s.
Here, the numerator and denominator compete and consequently $h_c(0)$
first increases before showing a slow decrease to its BCS value. Finally, $y$ in formula
(\ref{eq:yratio}) decreases with increasing offdiagonal $\lambda$ because
of the square bracket in the denominator. 
It is clear from these comparisons between Eliashberg and RBCS
that, in general, 
both strong coupling and anisotropy effects play a significant role in the
dimensionless ratios, and both need to be accounted for.

\begin{figure}[ht]
\begin{picture}(250,200)
\leavevmode\centering\includegraphics{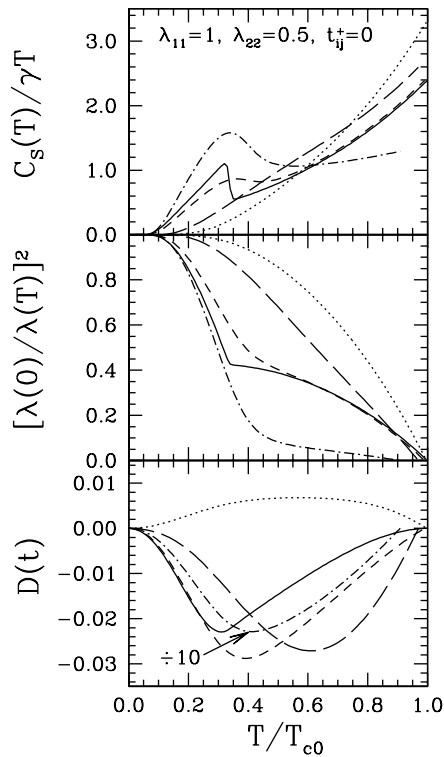}
\end{picture}
\vskip 70pt
%\begin{figure}
%\includegraphics[clip,width=0.45\textwidth]{tc1.eps}
%\includegraphics{tc1.eps}
\caption{Upper frame:
Specific heat in the superconducting state 
normalized to the normal state, $C_S(T)/\gamma T$, versus $T/T_{c0}$,
 where
$T_{c0}$ 
is the $T_c$ for only the $\lambda_{11}$ channel, with all
others zero. Shown are curves for various offdiagonal $\lambda$'s
with $\lambda_{11}=1$ and $\lambda_{22}=0.5$. Three curves are for
$\lambda_{12}=\lambda_{21}$ equal to: 0.0001 (solid), 0.01 (short-dashed)
and 0.1 (long-dashed). Also shown are: $\lambda_{12}=0.1$ and $\lambda_{21}
=0.01$ ({\it i.e.}, $\alpha =10$) (dot-dashed) and 
$\lambda_{12}=0.01$ and $\lambda_{21}=0.1$ ({\it i.e.}, $\alpha =0.1$) (dotted).
Middle frame: The superfluid density $[\lambda(0)/\lambda(T)]^2$ versus
$T/T_{c0}$ for the same parameters.
Lower frame: The deviation function $D(t)$ plotted  versus
$T/T_{c0}$. The dot-dashed curve has been divided by 10 from its original
value in order to display it on the same scale as the other curves.}
\label{fig9}
\end{figure}

\section{The limit of nearly separate bands}

When $\lambda_{12}=\lambda_{21}=0$, there exist two 
transition temperatures $T_{c1}$ and $T_{c2}$ associated with
$\lambda_{11}$ and $\lambda_{22}$, separately, and for several
properties, but not all, the 
superconducting state is the straight sum of the two bands as they would
be in isolation. Here, we wish to study how the integration of the
two bands proceeds as $\lambda_{12}$ and/or $\lambda_{21}$ is switched
on. Our first results related to this issue are shown in Fig.~\ref{fig9},
which has three frames. The top frame deals with the normalized specific
heat $C_S(T)/\gamma T$ as a function of temperature, the middle,
the normalized inverse square of the penetration depth $[\lambda_L(0)/\lambda_L(T)]^2$ and the bottom gives the critical field deviation function $D(t)$
of Eq.~(\ref{eq:devfn}). In all cases, we have used our Lorentzian model
for the spectral densities $\alpha^2_{ij}F(\omega)$ with $\lambda_{11}=1$
and $\lambda_{22}=0.5$ fixed for all curves. The solid curves are
for $\lambda_{12}=\lambda_{21}=0.0001$, short-dashed for 0.01, and long-dashed 
for 0.1. 
In the top two frames, the two separate transitions
are easily identified in the curves with solid line type. 
Because of the very small value of $\lambda_{12}=\lambda_{21}$,
the composite curve is obviously the summation
of two subsystems,
which are almost completely decoupled. 
However, already for $\lambda_{12}=\lambda_{21}=0.01$
which remains very small as compared with the value of 
$\lambda_{11}$ and even $\lambda_{22}$, the second transition (short-dashed
curve) becomes significantly smeared. The two subsystems have undergone
considerable integration. In particular, the second specific heat jump
is rounded, becoming more knee-like. Also, the sharp edge or kink in the solid 
curve for the superfluid density is gone in the short-dashed curve.
Thus, to observe clearly two distinct systems, the offdiagonal
$\lambda$'s  need to be very small. Once $\lambda_{12}=\lambda_{21}=0.1$ 
(long-dashed curve), the integration of the two subsystems is very considerable
if not complete. This does not mean, however, that superconducting properties
become identical to those for an equivalent one-band system. As long as
the $\alpha^2_{ij}F(\omega)$ are not all the same, there will be 
anisotropy and this will change properties as compared with 
isotropic Eliashberg one-band results. 
Note that in the solid Eliashberg curve of Fig.~\ref{fig6}, a point of
inflection remains, as commented on by Golubov {\it et al.}\cite{dolgovpen}.
In the case of the deviation function (lower frame), the solid curve
shows a sharp cusp which is related to the lower transition temperature of
the decoupled bands but not quite at that value as this function is composed 
from subtracting $1-(T/T_c)^2$ from $H_c(T)/H_c(0)$. However, two
distinct pieces of the curve exist and notably near
$T_c$ the curve has
 a very different curvature from what is normally encountered.
In particular, the temperature dependence of the solid curve is 
concave down at high temperature in contrast to the usual case of concave
up. As the bands are coupled through larger and larger
interband $\lambda$'s,
the curve moves to a shape more consistent with one-band behaviour.
However, the curve remains negative due to the anisotropy, while
usually strong coupling  would drive the curve positive with an
overall concave-down curvature\cite{carbotte}, which is illustrated
by the dotted curve for which the first band dominates, as we describe
below.

The other curves in these figures,
dot-dashed and dotted, are for $\alpha=10$ with $\lambda_{12}=0.1$ and
$\lambda_{21}=0.01$, and $\alpha=0.1$ with $\lambda_{12}=0.01$ and
$\lambda_{21}=0.1$, respectively. 
For $\alpha=10$, the second band with the smaller of the two diagonal
values of $\lambda$ has ten times the density of states as compared to 
band 1 with the larger $\lambda$ value. This large disparity in
density of states can have drastic effects on superconducting properties,
and further modify both the observed temperature dependence and the value
of the BCS ratios.
The second specific heat jump in the dash-dotted curve, although 
smeared, is quite large as compared with that in the solid or even
the dashed curve. Also, it is to be noted that beyond the temperature
of the lower maximum in $C_S(T)/\gamma T$, the curve shows only a very modest
increase, reflecting the low value of the electronic density of states in
band 1, and the ratio of the jump at $T_c$ to the normal state
is now quite reduced. 
The low density of states in band 1
 is also reflected in the low value of the penetration depth
curve (middle frame, dash-dotted curve) in the temperature region
above $T_{c2}$.
Finally, we note that while we have chosen a large value of $\alpha$ for
illustration here, MgB$_2$ has an $\alpha = 1.37$ which, by the above
arguments, would tend to accentuate the features due to the second band.

\begin{figure}[ht]
\begin{picture}(250,200)
\leavevmode\centering\includegraphics{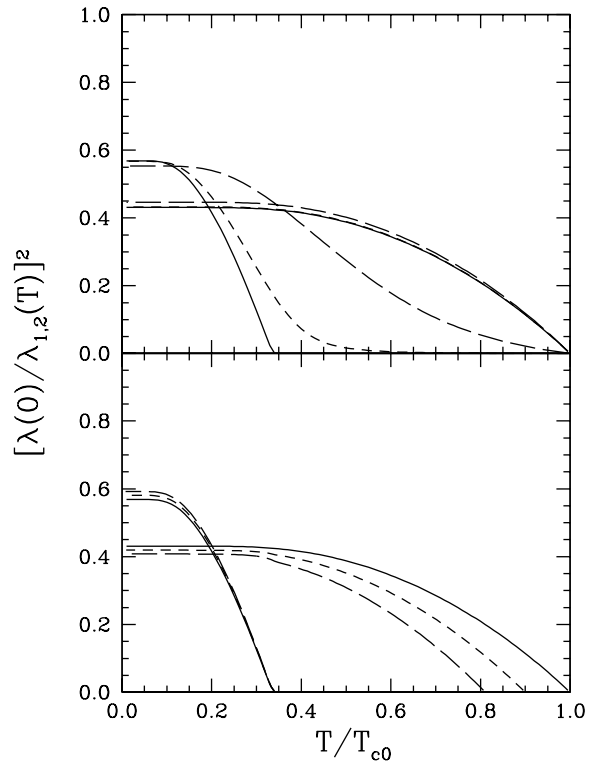}
\end{picture}
\vskip 70pt
%\begin{figure}
%\includegraphics[clip,width=0.45\textwidth]{tc1.eps}
%\includegraphics{tc1.eps}
\caption{Upper frame: Individual contributions from
each band to the superfluid density $[\lambda(0)/\lambda_{1,2}(T)]^2$
as a function of $T/T_{c0}$, where
$T_{c0}$ is the $T_c$ for the $\lambda_{11}$ channel alone, with all
others zero. Shown are curves for various offdiagonal $\lambda_{21}$
with $\lambda_{11}=1$, $\lambda_{22}=0.5$, and $\lambda_{12}=0.0001$. 
The three pairs of curves are for
$\lambda_{21}$ equal to: 0.0001 (solid), 0.01 (short-dashed),
and 0.1 (long-dashed). 
The curves which go to zero at a lower temperature correspond to
$[\lambda(0)/\lambda_{2}(T)]^2$ while those which go to zero close to 1
are for $[\lambda(0)/\lambda_{1}(T)]^2$.
Lower frame: Now the $\lambda_{21}$ is held fixed at 0.0001 and the
$\lambda_{12}$ is varied. 
The three pairs of curves are for
$\lambda_{12}$ equal to: 0.0001 (solid), 0.1 (short-dashed),
and 0.2 (long-dashed). Here,
the ratio of the density of states $\alpha$ has been taken to be 1
for convenience of illustrating the curves on the same scale.}
\label{fig10}
\end{figure}

A very different behaviour is obtained when $\alpha=0.1$ for which case
the electronic density of states in the second band is reduced by a factor
of ten as compared to the first band. In this case, the dotted curve applies 
and 
looks much more like a standard one-band case with very
significant strong coupling effects $\Delta C(T_c)/\gamma T_c\simeq 2.4$.
The influence of band 2 has been greatly reduced.  Finally, we note that
the introduction of the offdiagonal elements can change $T_c$. In
particular, the dot-dashed curve ends at a considerably reduced value of critical
temperature as compared with the other curves. This is consistent with
Fig.~\ref{fig1} where we saw that increasing $\lambda_{12}$ for small values of
$\lambda_{21}$ decreases $T_c$. On the other hand, for the dotted curve
for which values of $\lambda_{12}$ and $\lambda_{21}$ are interchanged
as compared to the dash-dotted curve, $T_c$ is hardly
affected because $\lambda_{12}$ is small and it is this parameter which
affects $T_c$ more. The two parameters $\lambda_{12}$ and $\lambda_{21}$
do not play the same role in $T_c$ or for that matter in the integration
process of the two bands. This is made clear in Fig.~\ref{fig10} which
deals only with the penetration depth. What is shown are the separate
contributions to the superfluid density coming from the two bands.
In all cases, $\lambda_{11}=1$ and $\lambda_{22}=0.5$. In the top frame,
$\lambda_{12}=0.0001$ and $\lambda_{21}$
is varied.
It is clear that as $\lambda_{21}$ is increased, the superfluid density
associated with the second band remains significant even above the second
transition temperature $T_{c2}$ which is well-defined in the solid curve.
This is the opposite behaviour of what is seen in the lower frame where
$\lambda_{21}$ remains at 0.0001 and $\lambda_{12}$ is increased.
In this case,
$T_c$ changes significantly but the superfluid density associated with
the second band remains negligible above $T_{c2}$. Note
finally that the relative 
size of the superfluid density in each band will vary with
$\alpha$ and $v_{Fi}$, neither of which have been properly accounted
for in this figure, as we wished to illustrate solely the effect of
$\lambda_{12}$ and $\lambda_{21}$ on the issue of integration of
the bands and modification of $T_c$.

\begin{figure}[ht]
\begin{picture}(250,200)
\leavevmode\centering\includegraphics{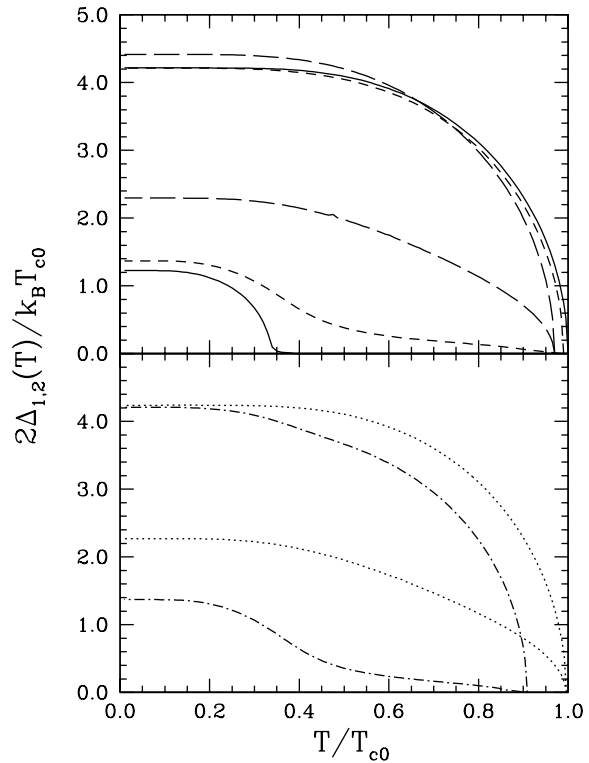}
\end{picture}
\vskip 70pt
%\begin{figure}
%\includegraphics[clip,width=0.45\textwidth]{tc1.eps}
%\includegraphics{tc1.eps}
\caption{Upper frame: Upper and lower gap ratios, $2\Delta_{1,2}/k_BT_{c0}$,
versus $T/T_{c0}$, where
$T_{c0}$ 
is the $T_c$ for the $\lambda_{11}$ channel alone, with all
$\lambda$'s zero. Shown are curves for various offdiagonal $\lambda$'s
with $\lambda_{11}=1$ and $\lambda_{22}=0.5$. Three pairs of curves are for
$\lambda_{12}=\lambda_{21}$ equal to: 0.0001 (solid), 0.01 (short-dashed)
and 0.1 (long-dashed). 
Lower frame: Same as for upper frame
except now are shown: $\lambda_{12}=0.1$ and $\lambda_{21}
=0.01$ ({\it i.e.}, $\alpha =10$) (dot-dashed) and 
$\lambda_{12}=0.01$ and $\lambda_{21}=0.1$ ({\it i.e.}, $\alpha =0.1$) (dotted).}
\label{fig11}
\end{figure}

The changes, with the offdiagonal elements $\lambda_{12}$ and $\lambda_{21}$,
in the temperature dependence of the upper and lower gaps are closely
correlated with those just described for the superfluid density.
This is documented in Fig.~\ref{fig11} which has two frames. In all
cases $\lambda_{11}=1$ and $\lambda_{22}=0.5$. In the top frame,
$\lambda_{12}=\lambda_{21}$ equal to 0.0001 (solid), 0.01 (short-dashed),
and 0.1 (long-dashed). The various pairs of curves apply to the 
upper and lower gap ratios. Note the long tails in the short-dashed
curve (lower gap), still small but extending to $T=T_c$. For the
long-dashed curve, the lower and upper gaps now have very similar
temperature dependences, but these
are not yet identical to
BCS. We have already seen in 
Fig.~\ref{fig7}, for the specific case of MgB$_2$, that the
lower gap falls below BCS at temperatures above $T/T_c\simeq 0.7$,
which is expected when it is viewed as an evolution
out of two separate gaps, with two
$T_c$ values, due to increasing the offdiagonal
coupling.
 In the lower frame, we show results for $\alpha=10$
(dot-dashed) and $\alpha=0.1$ (dotted). Again, as expected, the two
dash-dotted curves show distinct temperature dependences while for the
dotted they are very similar.

\begin{figure}[ht]
\begin{picture}(250,200)
\leavevmode\centering\includegraphics{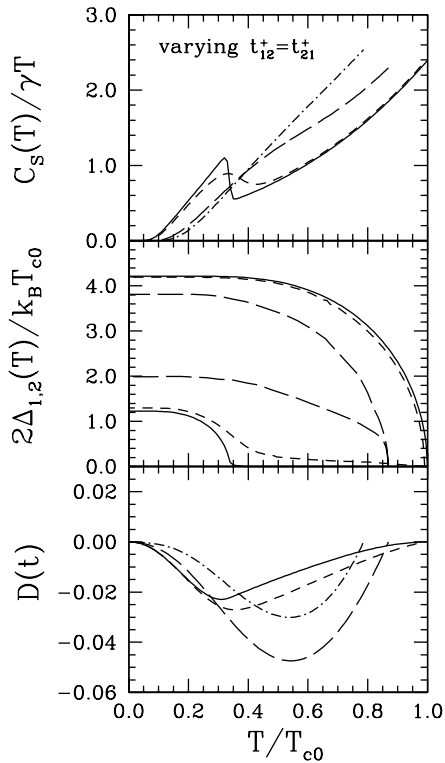}
\end{picture}
\vskip 70pt
%\begin{figure}
%\includegraphics[clip,width=0.45\textwidth]{tc1.eps}
%\includegraphics{tc1.eps}
\caption{Top frame: Specific heat in the superconducting state 
normalized to the normal state, $C_S(T)/\gamma T$, versus $T/T_{c0}$, where
$T_{c0}$ 
is the $T_c$ for the pure case. Here,
$\lambda_{11}=1$, $\lambda_{22}=0.5$, and $\lambda_{12}=\lambda_{21} = 0.0001$.
Shown are curves for varying $t^+_{12}=t^+_{21}$
 equal to: 0.0 (solid), 0.01 (short-dashed), 0.2
 (long-dashed), and 0.5 (dot-dashed) in units of $T_{c0}$. 
Notice that the value of the jump at $T_c$ first dips and then rises with
impurity scattering.
Middle frame: $2\Delta_{1,2}/k_BT_{c0}$ versus $T/T_{c0}$. The upper
three curves correspond to $2\Delta_1/k_BT_{c0}$ and the lower three
to $2\Delta_2/k_BT_{c0}$, with the curves labelled the same way as in
the upper frame. Only the first three impurity cases are shown
for clarity. The other progresses in the same manner with the $T_c$
reducing further and the gaps moving closer to a common value.
Bottom frame: The deviation function $D(t)$ versus $T/T_{c0}$,
again with the curves labelled the same way as in
the top frame.}
\label{fig12}
\end{figure}

\begin{figure}[ht]
\begin{picture}(250,200)
\leavevmode\centering\includegraphics{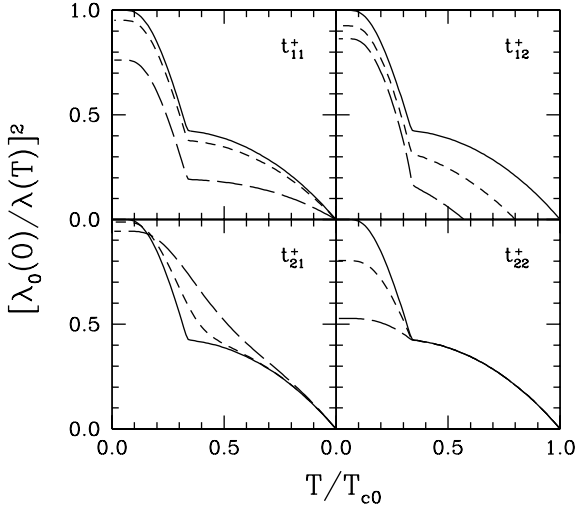}
\end{picture}
%\vskip 60pt
%\begin{figure}
%\includegraphics[clip,width=0.45\textwidth]{tc1.eps}
%\includegraphics{tc1.eps}
\caption{Effect of impurity scattering on the
 superfluid density $[\lambda_0(0)/\lambda(T)]^2$, plotted versus
$T/T_{c0}$,  where
$T_{c0}$ and $\lambda_0(0)$ are for the pure case.
Each frame shows the effect of the different type of impurity scattering
keeping all other impurity terms equal to zero. The spectrum parameters
are:
$\lambda_{11}=1$, $\lambda_{22}=0.5$, and
$\lambda_{12}=\lambda_{21} =0.0001$. The solid curve in
all cases is for the pure case of $t^+_{ij}=0$.
In the upper left frame: $t^+_{11}/T_{c0} = 0.2$ (short-dashed) and 2.0 (long-dashed).
In the upper right frame: $t^+_{12}/T_{c0} = 0.2$ (short-dashed) and 0.4 (long-dashed).
In the lower left frame: $t^+_{21}/T_{c0} = 0.02$ (short-dashed) and 0.1 (long-dashed).
In the lower right frame: $t^+_{22}/T_{c0} = 0.2$ (short-dashed) and 2.0 (long-dashed).
}
\label{fig13}
\end{figure}

A very similar story emerges when interband impurity scattering is
considered. Results are given in Fig.~\ref{fig12} and Fig.~\ref{fig13}.
Fig.~\ref{fig12} has three frames. Here, $\lambda_{11}=1$, $\lambda_{22}=0.5$, and $\lambda_{12}=\lambda_{21}=0.0001$,
  with our Lorentzian electron-phonon
spectral functions $\alpha^2_{ij}F(\omega)$ described previously. The top
frame deals with the temperature dependence of the normalized superconducting
state electronic specific heat $C_S(T)/\gamma T$. The middle frame gives the
gap ratios of $\Delta_1$ and $\Delta_2$ and thus the curves come in pairs,
with $\Delta_1>\Delta_2$. And the bottom
frame shows the deviation function $D(t)$ for the thermodynamic
critical
magnetic field.  What is varied in the various curves is the interband
impurity scattering rate $t^+_{12}=t^+_{21}$ (taken to be equal in value,
{\it i.e.}, $\alpha=1$). The solid curve, which clearly shows two transitions,
is for $t^+_{12}=0$.
It is
to be noted first, that in all cases, offdiagonal impurity scattering changes
the value of the critical temperature, reducing it to less than 0.8 of its
pure value in the case of the dot-dashed curve. This decrease in $T_c$ does not
translate, however, into a steady decrease in the specific heat jump at
$T_c$. We see that while the jump initially decreases with increasing
$t^+_{12}=t^+_{21}$, eventually it increases and is largest for the dot-dashed
curve. Both Watanabe and Kita\cite{kita} and
Mishonov {\it et al.}\cite{mishonov}, using only an unrenormalized
BCS model, find an increase with impurity scattering and no initial
decrease as is found in the full Eliashberg calculation. This is a clear
illustration that, at minimum, a renormalized BCS formula needs to be used
to capture the qualitative trend and full Eliashberg theory is
required if  one wishes to
be quantitative.
It is also clear that as interband impurity scattering increases,
the jump in the specific heat at the second transition, seen in the solid 
curve, is rapidly washed out and little remains of this anomaly in the
dot-dashed curve. Even the long-dashed curve
shows little structure in this region, in analogy
to what we found to hold for the case of increasing the 
offdiagonal electron-phonon
elements. Note, however, there remains 
a point of inflection which has moved to higher temperature.
Such a shift of the inflection point 
can also be brought about by increasing the 
offdiagonal $\lambda$'s as seen in Fig.~\ref{fig9}.

The temperature dependence of the
gap ratios (middle frame of Fig.~\ref{fig12}) also mirror what we found 
in Fig.~\ref{fig11}. The dashed curves exhibit quite distinct temperature
dependences between $\Delta_1$ and $\Delta_2$ while this is no longer
the case for the pair of long-dashed curves. Note that, as compared to the solid curve, the anisotropy in the gaps for the long-dashed curve has been reduced
considerably. The upper gap has decreased and the lower increased even more.
The washing out of the gap anisotropy by offdiagonal impurity scattering
is expected and has been studied theoretically\cite{golubovimp,erwin}
and experimentally\cite{gonnelli04}. For carbon doping, the gaps
are seen to merge at about 13\% for which $T_c$ has
been reduced to about 20 K with the large gap reducing to its BCS value
and the smaller gap moving upwards only very little in contrast
to our model calculations for which the lower gap changes relatively
more and isotropy is reached at about a 30\% reduction in $T_c$. Of course, as
one dopes, the electronic density of states and the electron-phonon
parameters also change\cite{ummarino}, and one needs to include these
in addition to any interband scattering.

Finally, the effect of interband ordinary impurity scattering on the 
deviation function, shows a behaviour similar to that found for
paramagnetic impurities in one-band superconductors\cite{ewald}.
Initially, as in the other properties, the impurities smear the
structure related to the second transition temperature (in this case the
cusp feature in the solid curve) and once
the two bands are fairly well integrated,
then like paramagnetic impurities, the effect here is to keep the
minimum at the same temperature but change its value. A key difference
though is that, in the case of paramagnetic impurities in one-band, the
 extremum in the curve moves from positive (and strong coupling)
to negative (and weak coupling) because the gap is being reduced towards
zero. Here, with the two bands, the impurities do not reduce $T_c$, and
hence the gap, to zero, but rather to a finite value related to the washing
out of the anisotropy between the two bands, and hence the extremum in this case
will move from negative (where it is positioned due to large anisotropy)
to smaller values, reflecting this.

Next we turn to the results given in Fig.~\ref{fig13}, which shows the temperature
dependence of the superfluid density for various values of impurity parameters.
Again, $\lambda_{11}=1$, $\lambda_{22}=0.5$, and $\lambda_{12}=\lambda_{21}=0.0001$  with the Lorentzian spectra. What is illustrated in these four frames
is how very different the effect of $t^+_{11}$, $t^+_{22}$, $t^+_{12}$,
and $t^+_{21}$ are. The solid curve is for reference and is the pure case.
Once again, for the case of varying $t^+_{12}$ and $t^+_{21}$, we have
violated the constraint that their ratio must be fixed by the ratio of the
density of states. This we can do theoretically to decouple
and, therefore, illustrate
the effects of these different scattering channels, but in real systems,
they would be constrained and the net result would be a combination of the
effects from both channels.
The top left frame shows the effect on the superfluid density of increasing the
impurity scattering in the first band (intraband scattering). 
Such impurities reduce the superfluid density in band 1 while leaving band
2 unaltered. In the lower right-hand
frame it is the superfluid density in the second band that is reduced, leaving
the first unchanged.
$T_c$ is unaffected by intraband impurity scattering in
isotropic s-wave superconductors due to Anderson's theorem. 
The top right-hand frame shows that increasing $t^+_{12}$ reduces the
critical temperature as well as reduces the superfluid density in band
one without, however, having much effect on the second band. The kink 
associated with the rise of the second band is hardly changed as
$t^+_{12}$ is not the integrating variable, rather it is $t^+_{21}$ which
integrates rapidly the bands as seen in the lower left-hand frame.
However, in this case, the critical temperature is hardly changed
and there is little change to the curve above $T/T_{c0}\simeq 0.7$. 

\section{The limit of pure offdiagonal coupling}

While the two-band nature  in
MgB$_2$, driven by the electron-phonon interaction,
is well-established, there have been many reports of possible
two-band superconductivity in other systems, including the
conventional A15 compound Nb$_3$Sn\cite{guritanu}. With a $T_c=18$ K
and a main gap $2\Delta_M\sim 4.9T_c$, there is specific heat evidence
for a second gap at $0.8T_c$. Other systems are NbSe$_2$\cite{boaknin},
Y$_2$C$_3$ and La$_2$C$_3$\cite{sergienko} and possibly
a second nonsuperconducting band in CeCoIn$_5$\cite{tanatar}.
In the triplet spin state superconductor
Sr$_2$RuO$_4$\cite{mackenzie}, a small gap is induced in the second
band. As two-band superconductivity is likely to be a widespread
phenomenon, not confined to electron-phonon systems, it seems 
appropriate to investigate further an extended range of parameter
space for the $\lambda_{ij}$'s and in particular the
possibility that the offdiagonal elements are the dominant mechanism
for superconductivity.

\begin{figure}[ht]
\begin{picture}(250,200)
\leavevmode\centering\includegraphics{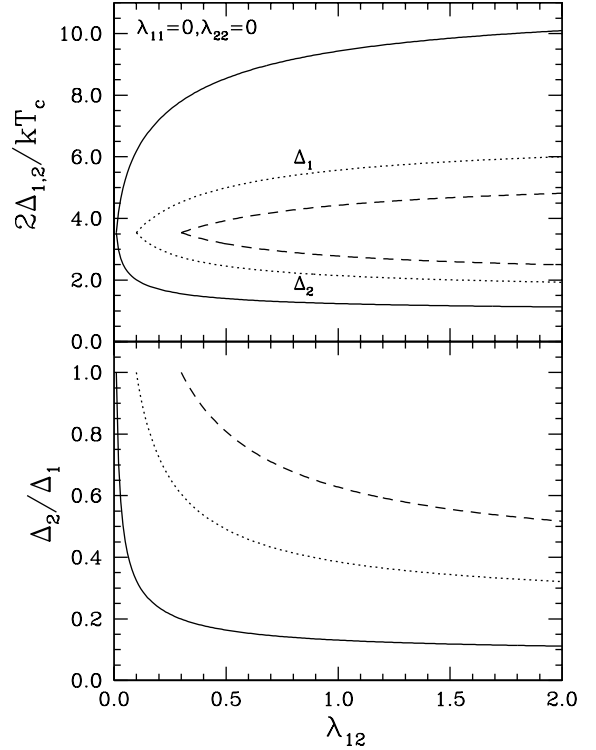}
\end{picture}
\vskip 70pt
%\begin{figure}
%\includegraphics[clip,width=0.45\textwidth]{tc1.eps}
%\includegraphics{tc1.eps}
\caption{Upper frame: Gap ratios for the upper ($2\Delta_1/k_BT_c$)
and lower gap ($2\Delta_2/k_BT_c$) as a function of $\lambda_{12}$
for varying $\lambda_{21}$: 0.01 (solid), 0.1 (dotted),
and 0.3 (dashed). Here, $\lambda_{11}=0$, $\lambda_{22}=0$.
This is for comparison with Suhl {\it et al.}\cite{suhl}.
Lower frame: Gap anisotropy, $u=\Delta_2/\Delta_1$, versus $\lambda_{12}$ for the
same parameters and curve labels as the upper frame. Note that $\lambda_{12}\ge
\lambda_{21}$ is plotted. With $\lambda_{12}<\lambda_{21}$, 
the roles
are simply reversed with $1\leftrightarrow 2$ and $\Delta_2$ would become
the large gap, etc.} 
\label{fig14}
\end{figure}

In the limit of pure offdiagonal coupling,
where $\lambda_{11}=\lambda_{22}=0$, Eq.~(\ref{eq:A}) for the
coupling $A$, which determines $T_c$ from Eq.~(\ref{eq:Tc}), simplifies
to:
\begin{equation}
A=\frac{1}{\sqrt{\bar\lambda_{12}\bar\lambda_{21}}},
\end{equation}
and the ratio of the gap to $T_c$ given in Eq.~(\ref{eq:gratio}) becomes
\begin{equation}
\frac{2\Delta_1}{k_BT_c}=3.54 {\rm exp}\biggl[A-\frac{u}{\bar\lambda_{21}}\biggr].
\end{equation}
The ratio $\bar\lambda_{12}/\bar\lambda_{21}=\alpha^*$ can be taken
$\ge 1$ and Eq.~(\ref{eq:u}) for the gap anisotropy $u=\Delta_2/\Delta_1$ written as:
\begin{equation}
\alpha^*u^2-1=\frac{\sqrt{\alpha^*}}{A}u\ln u.
\label{eq:odu}
\end{equation}
This equation gives $u$ in terms of $\alpha^*$ and $A$. Since by its
definition $0<u\le 1$, $u\ln u$ is negative so a condition on obtaining a
solution of (\ref{eq:odu}) is that 
\begin{equation}
\alpha^*u^2-1< 0 \,\,\, {\rm or}\,\,\, \alpha^*\le \frac{1}{u^2}.
\end{equation}
For a trial solution of $u=0.1$, this would give $1<\alpha^*<100$.
For $\alpha^*=60$, as an example, $A=4.46$ and $2\Delta_1/k_BT_c\simeq 9.7$,
which is very large. This occurs for $T_c/\omega_{\ln}\sim 10^{-2}$,
using $\ln(1.13\omega_{\ln}/T_c)=A$, which is in the weak coupling
regime. However, to achieve an upper gap ratio value greater than 11 or so,
will correspond an unrealistically small value of $T_c/\omega_{\ln}$
(of order $10^{-10}$, for example). 
In Fig.~\ref{fig14}, 
we show results in the upper frame
for $2\Delta_{1,2}/k_BT_c$ versus $\lambda_{12}$
for various $\lambda_{21}$ values.
 In the lower frame, we show $u$ versus $\lambda_{12}$.
The difference between Fig.~\ref{fig14} and Fig.~\ref{fig3}
shows that large values of $2\Delta_1/k_BT_c$ are more naturally obtained
in the pure offdiagonal regime and are associated as well with small
values of $u$ and the weak coupling regime. 
This latter feature implies that there will be no further strong
coupling corrections to an already large gap ratio.
We have also calculated
the thermodynamics and superfluid density in this regime, for a range
of parameters, but have found these properties to show quite ordinary
behaviour and have discovered  no new physics. For the sake of brevity,
we present none of these results but instead note that in this limit
the $\chi$'s are
\begin{eqnarray}
\chi_1 &=&\frac{1}{2}\biggl[1+\frac{\bar\lambda_{21}}{\bar\lambda_{12}}\biggr]
=\frac{1}{2}\biggl[1+\frac{1}{\alpha^*}\biggr],\\
\chi_2 &=&\frac{1}{2}\biggl[1+\frac{\bar\lambda_{12}}{\bar\lambda_{21}}\biggr]
 = \frac{1}{2}(1+\alpha^*),
\end{eqnarray}
with $\alpha^*=\bar\lambda_{12}/\bar\lambda_{21}$,
and hence, the various dimensionless ratios are:
\begin{equation}
\frac{\Delta C}{\gamma T_c}=1.43\biggl[\frac{4\alpha^*}{(1+\alpha^*)^2}\biggr]
\end{equation}
and
\begin{equation}
h_c(0)= \frac{\Delta_1}{T_c}\frac{1}{\pi}\sqrt{\frac{7\zeta(3)}{32}}
(1+\alpha^*)\sqrt{\frac{1+\alpha u^2}{\alpha+\alpha^{*2}}}
\end{equation}
and
\begin{equation}
\frac{y_L(0)}{|y_L'(T_c)|T_c}=\frac{1}{4}\frac{(1+\alpha\beta)(1+\alpha^*)}
{\alpha^*+\beta\alpha},
\end{equation}
where $\beta=v^2_{F2}(1+\lambda_{12})/v^2_{F1}(1+\lambda_{21})$.
The ratio for the zero temperature critical field of Eq.~(\ref{eq:hratio})
does not change its form and so is not repeated here.
These ratios
 behave, qualitatively, no differently from what we found in
section III. A difference worth noting is the following. In linear
order, the effect  of interband impurity scattering on $T_c$ takes the
form (\ref{eq:tcimp}-\ref{eq:rhoimp}):
\begin{equation}
\frac{\Delta T_c}{T_{c0}}=-\frac{\pi^2}{8}\rho^\pm_{12}
\biggl[1\mp \sqrt{\frac{\bar\lambda_{21}}{\bar\lambda_{12}}}\biggr]^2
\label{eq:odtcimp}
\end{equation}
which is always negative and larger for paramagnetic
than for normal impurities. It can also be very large
for $\rho^\pm_{12}\gg 1$.
This is another distinction between pure offdiagonal coupling and MgB$_2$,
for example. In obtaining (\ref{eq:odtcimp}), we have used the fact that
$\bar\lambda_{12}/\bar\lambda_{21}=\rho_{12}/\rho_{21}$ and $\bar\lambda_{12}/\bar\lambda_{21}>1$.

\section{Conclusions}

We have calculated thermodynamics, gap anisotropy
and penetration depth for a two-band Eliashberg superconductor.
For the parameters appropriate to MgB$_2$ which are obtained from
first principle band structure calculations of the
electron-phonon spectral functions, we find
good agreement with the existing experimental data. We reduce
the Eliashberg equations to a renormalized BCS form by application
of the two-square-well approximation. Comparison of these results
with those from the full Eliashberg equations allows us to determine
strong coupling corrections, which we find to be significant in
MgB$_2$. When the parameters for the electron-phonon interaction
are moved away from those specific to MgB$_2$, the strong coupling
corrections can become much larger, and superconducting properties
reflect this fact, as well as the change in anisotropy between the bands.
Within the $\lambda^{\theta\theta}$ approximation, we derive simple
analytic expressions for the various dimensionless BCS ratios which
would be universal in the one-band case, but are not in the two-band one.
They depend on the anisotropy and particularly on the ratio of the
electronic density of states in the two bands. The anisotropy
in the ratio of the two gaps at zero temperature is investigated and
is found to increase as $\lambda_{22}$ is reduced and made repulsive,
in which case the existence of superconductivity in the first band,
and the offdiagonal coupling to it, induces a gap
in a band which would, on its own, not be superconducting.

We have paid particular attention to the limit of nearly decoupled
bands, {\it i.e.}, small interband coupling, with the superconductivity
originating from $\lambda_{11}$ and $\lambda_{22}$ in the first and
second band, respectively. When $\lambda_{12}, \lambda_{21}\to 0$, there are
two transitions at $T_{c1}$ and $T_{c2}$ and two specific
heat jumps. As the interband coupling is turned on, the
two bands become integrated and the second transition smears. We have
found that the two parameters, $\lambda_{12}$ and $\lambda_{21}$,
have very different effects on the smearing of the second transition
and on $T_c$. $\lambda_{12}$ largely modifies $T_c$, reducing it,
whereas, $\lambda_{21}$ alters the lower temperature region around
the second transition.
Only very small values of $\lambda_{21}$, as compared 
with $\lambda_{11}$ and $\lambda_{22}$, are needed to cause large changes
in the region around $T_{c2}$. It was found that a small
amount of interband impurity scattering can also significantly smear
the second transition,
and so reduce the distinction between the two bands. However, even when the
two bands are well-integrated and a sharp second transition 
is no longer easily discernible,
this does not imply that the superconducting properties become those of a
one-band superconductor. Anisotropy remains and this affects
properties.

In view of the possible widespread occurrence of two-band superconductivity,
even for systems with exotic mechanism not necessarily due to the
electron-phonon interaction, we deemed it of interest to consider 
the case of zero intraband coupling,
 $\lambda_{11}=\lambda_{22}=0$,
with superconductivity due only to the interband $\lambda_{12}$ and $\lambda_{21}$,
 which need not have  the same value. When these are very different, the
resulting gaps are quite different from each other and the ratio of 
$\Delta_1$ to $T_c$ can become large particularly in the weak
coupling limit. This is a distinguishing feature of pure offdiagonal coupling.
Another distinguishing feature is the possibility of a
rapid reduction of $T_c$ towards zero
by interband impurity scattering, as compared with the case for which
the diagonal elements play the leading role.

\begin{acknowledgments}
EJN acknowledges funding from NSERC, the
Government of Ontario
(Premier's Research Excellence Award), and the University of Guelph.
JPC acknowledges support from NSERC and the CIAR. 
In addition, we thank 
E. Schachinger and J. Wei for
helpful discussions.
\end{acknowledgments}


\begin{thebibliography}{99}
\bibitem{mustar} Note that in implementing the $\lambda^{\theta\theta}$
model with a finite $\mu_{ij}^*$, in principle, one cannot use the 
same $\mu_{ij}^*$ as those
used in the Eliashberg calculation as the Eliashberg $\mu_{ij}^*$
depend on the high frequency cutoff $\omega_c$, which is 750 meV in
our calculations. The $\lambda^{\theta\theta}$ model is based on
modeling the frequency dependent gap function by a constant up to
some cutoff which is on the scale of the phonon energies. Typically,
an appropriate cutoff might be the maximum phonon frequency or
the frequency beyond which the phonon spectrum suddenly drops to small
values, as this will be reflected in the gap function by a sudden drop
with the gap function going negative. Thus, to maximize the effectiveness
of this approximation to mimic the behaviour of the
gap function, one might use $\omega_\circ=75$ meV (as has 
Mitrovi\'c\cite{bozaimp}) as a major peak in the phonon spectra occurs
at this energy, after which it drops to small values. The new $\mu^*_{ij}(\omega_\circ)$ for the RBCS calculations would
 then found from the matrix equation: $\hat\mu^*(\omega_\circ)
=[1+\hat\mu^*(\omega_c)\ln(\omega_c/\omega_\circ)]^{-1}\hat\mu^*(\omega_c)$
\cite{bozaimp}.
For simplicity, we have used the Eliashberg $\mu^*$'s
in our calculations for Table~\ref{table1},
as we have found that for varying $\mu^*$, the values do not
change very much and hardly at all in some cases, and our goal is only
to provide a rough estimate of the strong coupling corrections in relation
to RBCS.
\end{thebibliography}
\end{document}